\documentstyle[12pt,aaspp4,flushrt]{article}
\def\approxgt{\lower.2em\hbox{$\buildrel > \over \sim$}}
\def\approxlt{\lower.2em\hbox{$\buildrel < \over \sim$}}
\def\cii{{\rm C}\thinspace{\sc{ii}}}
\def\civ{{\rm C}\thinspace{\sc{iv}}}
\def\hi{{\rm H}\thinspace{\sc{i}}}

\def\heii{{\rm He}\thinspace{\sc{ii}}}
\def\siliii{{\rm Si}\thinspace{\sc{iii}}}
\def\siliv{{\rm Si}\thinspace{\sc{iv}}}

\def\eg{{\it e.g.\/}}
\def\etal{{\it et~al.\/}}
\newcount\hours
 \newcount\minutes
  \def\SetTime{\hours=\time
         \global\divide\hours by 60
         \minutes=\hours
         \multiply\minutes by 60
         \advance\minutes by-\time
         \global\multiply\minutes by-1 }
 \SetTime
 \def\now{\number\hours:\ifnum\minutes<10 0\fi\number\minutes}

\begin{document}
\slugcomment{To appear in {\it The Astronomical Journal}} 
\title{The Redshift Evolution of the Ly$\alpha$ Forest}

\author{Tae-Sun Kim\altaffilmark{1}, Esther M. Hu\altaffilmark{1}, 
  Lennox L. Cowie\altaffilmark{1}, and Antoinette Songaila\altaffilmark{1}}
\affil{Institute for Astronomy, University of Hawaii, 2680 Woodlawn Dr.,
  Honolulu, HI 96822\\
  tsk@ifa.hawaii.edu, hu@ifa.hawaii.edu, cowie@ifa.hawaii.edu, 
  acowie@ifa.hawaii.edu}
\altaffiltext{1}{Visiting Astronomer, W. M. Keck Observatory, jointly
  operated by the California Institute of Technology and the University of
  California.}

\begin{abstract}
We have analyzed the properties of low column density Ly$\alpha$ forest
clouds (12.8 $\le \log N_{\rm HI} \le$ 16.0) toward 5 QSOs at
different redshifts, using high signal-to-noise data from the HIRES
spectrograph on the Keck{\thinspace}I 10-m telescope. The results are
used to study the evolution of these clouds in the interval $2.1 < z <
3.5$.  We find:  1) The differential density distribution of forest
clouds, $f(N_{\rm HI})$, fits an empirical power law with a typical
slope of $-1.35 \to -1.55$ for clouds with $N_{\rm HI} \le
10^{14.3}\ {\rm cm}^{-2}$ and changes extremely slowly with redshift
over this $N_{\rm HI}$ range. 2) The deviation of $f(N_{\rm HI})$ from
this power-law distribution at higher column densities depends on
redshift: for higher redshifts, the departure from a power-law
distribution occurs at higher $N_{\rm HI}$, which means that higher
column density clouds rapidly disappear with decreasing redshift.  This
may be consistent with an invariant shape for the different column
density distributions, but with a shift to a lower column density
normalization for systems at different redshifts driven by the overall
expansion of the universe.  3) The line widths of the forest clouds
increase as redshift decreases.  4) The correlation strength of the
forest clouds seems to increase as redshift decreases.  We discuss these
results in terms of the evolution of the IGM comparing the results to
models and analytic descriptions of the evolution of structure in the
gas.
\end{abstract}


\section{Introduction}

Ly$\alpha$ forest clouds imprinted in the spectra of high-redshift QSOs
provide a unique opportunity to study the evolution of cosmic structure
and the distribution of matter in the universe up to $z \sim 5$. The low
column density end of the forest is of particular interest, since the
number density of forest clouds increases as $N_{\rm HI}$ decreases down
to the current observational limit, $N_{\rm HI} = 3 \times 10^{12}\ {\rm
cm}^{-2}$ (\markcite{hu95}Hu \etal\ 1995, \markcite{lu96}Lu \etal\ 1996,
\markcite{kir97}Kirkman \& Tytler 1997), and the internal density in
these clouds is thought to lie close to the mean baryon density in the
intergalactic medium (IGM).  Although much observational progress has
been made since the advent of the Keck{\thinspace}I 10m telescope (which
has provided much more sensitive observations of the high-$z\/$ clouds)
and HST (which has accessed the low-$z\/$ clouds), the evolution of the
low column density clouds with redshift has not yet been described. The
reason for this is the need for time-consuming high-resolution
spectroscopic observations which are essential to detect the forest
clouds with \hi\ column densities less than $10^{14}\ {\rm cm}^{-2}$.
It is this topic which we address in the present paper.

Recently, a number of CDM-based numerical simulations 
have shown that Ly$\alpha$ forest clouds are a natural
consequence of structure formation in the intergalactic gas
(\markcite{cen94}Cen \etal\ 1994, \markcite{zha95}Zhang \etal\ 1995,
\markcite{her96}Hernquist \etal\ 1996, \markcite{dav97}Dav\'e
\etal\ 1997, \markcite{zha97}Zhang \etal\ 1997).  Although the
detailed simulation results and input cosmological parameters vary
from model to model, there are general trends which may be understood 
in broad analytic terms (\eg, \markcite{hui96}Hui \etal\ 1996, 
\markcite{bi97}Bi \& Davidsen 1997).  In particular, the
low column density forest ($N_{\rm HI} \le 10^{14}\ {\rm cm}^{-2}$)
forms in regions where the deviations from the mean density are
still small, and shocking has yet to occur.  (The lowest column
density clouds may arise in regions with density lower than the
mean density.)  The normalization of the column density distribution
in this range depends on the ionizing flux, the mean baryon
density, and the ionization history of the gas, but the slope of
this distribution depends primarily on the amount of power in the
fluctuation spectrum at near-Mpc scales, and weakly on the thermal
history.  Thus the evolution of the distribution function in column
density of these clouds is a powerful diagnostic of the formation of
structure in the IGM.

Here we use high-resolution, high signal-to-noise spectroscopic data
obtained from the Keck{\thinspace}I 10m telescope to study the physical
properties of lower column density ($N_{\rm HI} \le 10^{16}\ {\rm
cm}^{-2}$) forest clouds at $z > 2$ in three different redshift bins.
In particular, we address the following three questions: 1) Is there
any change in the number density of the lowest column density forest
clouds with redshift? 2) Is there any change in the broadening of the
clouds with redshift? 3) Is there any significant change in clustering
properties with redshift?  Sec.\ \ref{sec:observ} describes the
observations and data analysis.  The observational results are given in
Sec.\ \ref{sec:results}. In Sec.\ \ref{sec:discuss} we discuss the
observations from the perspective of the numerical simulations and also
in a broader context.  We summarize our conclusions in
Sec.\ \ref{sec:summary}.

\section{Observations and Data Analysis\label{sec:observ}}

The spectra of five QSOs (Q1422+231, Q0014+813, Q0302--003, Q1700+643,
and Q1623+268) suitable for studying the Ly$\alpha$ forest at $z > 2$
were selected from an on-going program to study QSO absorption line
systems (\markcite{son97}Songaila 1997). The properties of the clouds
seen toward Q0014+813 and Q0302--003 have already been described in
\markcite{hu95}Hu \etal\ (1995), but the current analysis includes
subsequently obtained data which substantially increased the exposure
times.  All QSOs were observed at the Keck{\thinspace}I 10m telescope
with the HIRES spectrograph at a resolution, $R$, of 36,000 ({\it cf}.
\markcite{cow95}Cowie \etal\ 1995, \markcite{hu95}Hu \etal\ 1995).  The
properties of the quasars and the exposure times are summarized in
Table \ref{tbl-1}.

The wavelength ranges used for this study were chosen to exclude any
higher order hydrogen Lyman absorption lines beyond Ly$\alpha$ and to
avoid the proximity effect. Two of the QSOs, Q1422+231 and Q0014+813,
have partial Lyman limit systems within the selected wavelength
ranges.  Ly$\alpha$ forest clouds within $\pm 1000~{\rm km\ s}^{-1}$
of the partial Lyman limit systems in both QSOs were excluded in order
to sample only the ``quiet'' forest (without possible influence
from the partial Lyman limit systems).  Table \ref{tbl-1} also gives 
details of the wavelength range and cloud properties.

Voigt profiles, which at these low \hi\ column densities are Gaussians,
were fitted to the absorption lines using an automatic line fitting
program (\markcite{hu95}Hu \etal\ 1995).  Because the lines are formed by
a combination of kinematic and thermal broadening, the profile fitting
method should not be viewed as a direct model of the underlying reality,
but rather as an extremely powerful parameterization of the spectrum
which can be compared to similar analyses of simulated model spectra
(\eg, \markcite{dav97}Dav\'e \etal\ 1997, \markcite{zha97}Zhang
\etal\ 1997) to test the models.  The present fitting program selects
lines with a central optical depth, $\tau \ge 0.05$, corresponding to an
\hi\ column density, $N_{\rm HI} = 2 \times 10^{12}\ {\rm cm}^{-2}$ at $b
= 30~{\rm km\ s}^{-1}$ (for thermal broadening the Doppler parameter, $b
\equiv \sqrt{2kT/m_{p}}$, where $T$ is the temperature for a thermally
broadened line, $m_{p}$ is the proton mass, and $k$ is the Boltzmann
constant), and then fits the spectrum with an ensemble of profiles
characterized by the \hi\ column density, the line width $b$, and the
central redshift $z$. The program determines a minimal ensemble of lines,
since lines can always be fit by a larger number of narrower components.
This ambiguity can result in significant differences and may be the
primary cause of the differences between \markcite{kir97}Kirkman \&
Tytler (1997) and \markcite{hu95}Hu \etal\ (1995), who find somewhat
different $b$ value distributions at the same redshift.  The present
analysis is self-consistent in that the five quasars are examined using
the same techniques, and therefore the measured evolution should be
robust.  However, we discuss in Sec.\ \ref{sec:alt} alternative
techniques which show the basic results in other forms.  Our evolution
also matches well to the results of \markcite{lu96}Lu \etal\ (1996) at
higher $z$, despite the possible differences in analysis techniques.

All QSO spectra, except for that of Q1623+268, have S/N of $\approxgt40$
per resolution element, and Q1422+231 has S/N of near 140 per resolution
element.  In the case of Q1623+268 the low blue sensitivity of the HIRES
CCD at wavelengths shorter than 4000${\rm \AA}$ dropped the S/N to $\sim
20$ over the wavelengths studied.  To construct a homogeneous data set,
we only used the lines with $N_{\rm HI} \ge 6.3 \times 10^{12}\ {\rm
cm}^{-2}$ from the 4 other QSO spectra. This also restricts us to the
column density range where any incompleteness corrections are small
(\markcite{hu95}Hu \etal\ 1995).  Although it is hard to quantify the
saturation of the lines because of the spread in $b$ values, we begin to
see these effects at about $N_{\rm HI} \ge  1.5 \times 10^{14} \ {\rm
cm}^{-2}$ for $b = 30~{\rm km\ s}^{-1}$ if we define a saturated line as
a line with a residual flux $r_{\nu} = \exp(-\tau_{\nu})$ less than
0.025.  As long as the lines did not show any severe departures from a
Voigt profile, all saturated lines were fitted with a single component.

All possible metal-line candidates were examined.  Unlike the case
for the $z < 1$ forest, it is not straightforward to distinguish
metal lines in the forest at $z > 2$ due to the higher number density
of forest clouds. Keeping in mind that most metal lines have lower
$b$ values ($< 15~{\rm km\ s}^{-1}$) than those of forest clouds when
fitted as \hi\ lines (\markcite{hu95}Hu \etal\ 1995), it is rather easy
to recognize \civ\ ($\lambda\lambda 1548, 1550$) and \siliv\
($\lambda\lambda 1393, 1402$).  With the exception of the spectrum of
Q1700+643, most \civ\ and \siliv\ doublets commonly associated with higher
column density forest clouds either lie outside the wavelength
regions under consideration or were not found. Metal lines from the
lower redshift metal line systems were also examined and excluded.
However, broad metal lines such as \siliii\ ($\lambda1206$) are hard to
identify correctly in the forest and may not be completely removed.
In any case, the fraction of metal lines in the forest will not be
higher than 5\%, and the number counts are dominated by Poisson noise.

For Q1700+643, which is known to have at least 10 low-redshift metal line
systems (\markcite{rei93}Reimers \& Vogel 1993, \markcite{vog95}Vogel \&
Reimers 1995, \markcite{rod95}Rodriguez-Pascual \etal\ 1995), a thorough
search was made for metal lines and these were removed.  However, there
are 9 unidentified lines with $b \le 20~{\rm km\ s}^{-1}$ in the
Q1700+643 forest compared to 1 line in the Q1623+269 forest over the
column density range, $N_{\rm HI} = 10^{13.1} - 10^{15}\ {\rm cm}^{-2}$.
We included these unidentified lines in our list of Ly$\alpha$ forest
clouds; however, they are most likely unidentified metal lines from the
metal line systems towards Q1700+643.  Since the number of unidentified
narrow lines is at most 11\% of the $b \le 20~{\rm km\ s}^{-1}$ systems,
the inclusion or exclusion of these lines does not affect the conclusions
reached in this analysis.

We grouped the 5 QSOs into three different redshift bins (Table \ref{tbl-1}),
with central values of $<z> = 2.31$ (Q1623+268, Q1700+643), $<z> =
2.85$ (Q0014+813, Q0302--003), and $<z> = 3.35$ (Q1422+231). 

\section{Results\label{sec:results}}

\subsection{Differential Density Distribution Function} 

The differential density distribution function (hereafter DDF or
$f(N_{\rm HI})$) is defined as the number of absorbing systems per
unit redshift path per unit column density as a function of $N_{\rm
HI}$.  The redshift path $X(z)$ is defined by $X(z) \equiv {1\over 2}
[(1+z)^{2} -1]$ for $q_{\rm o} = 0$ or by $X(z) \equiv {2\over 3}
[(1+z)^{3/2} -1]$ for $q_{\rm o} = 0.5$.  

\begin{figure}
\plotone{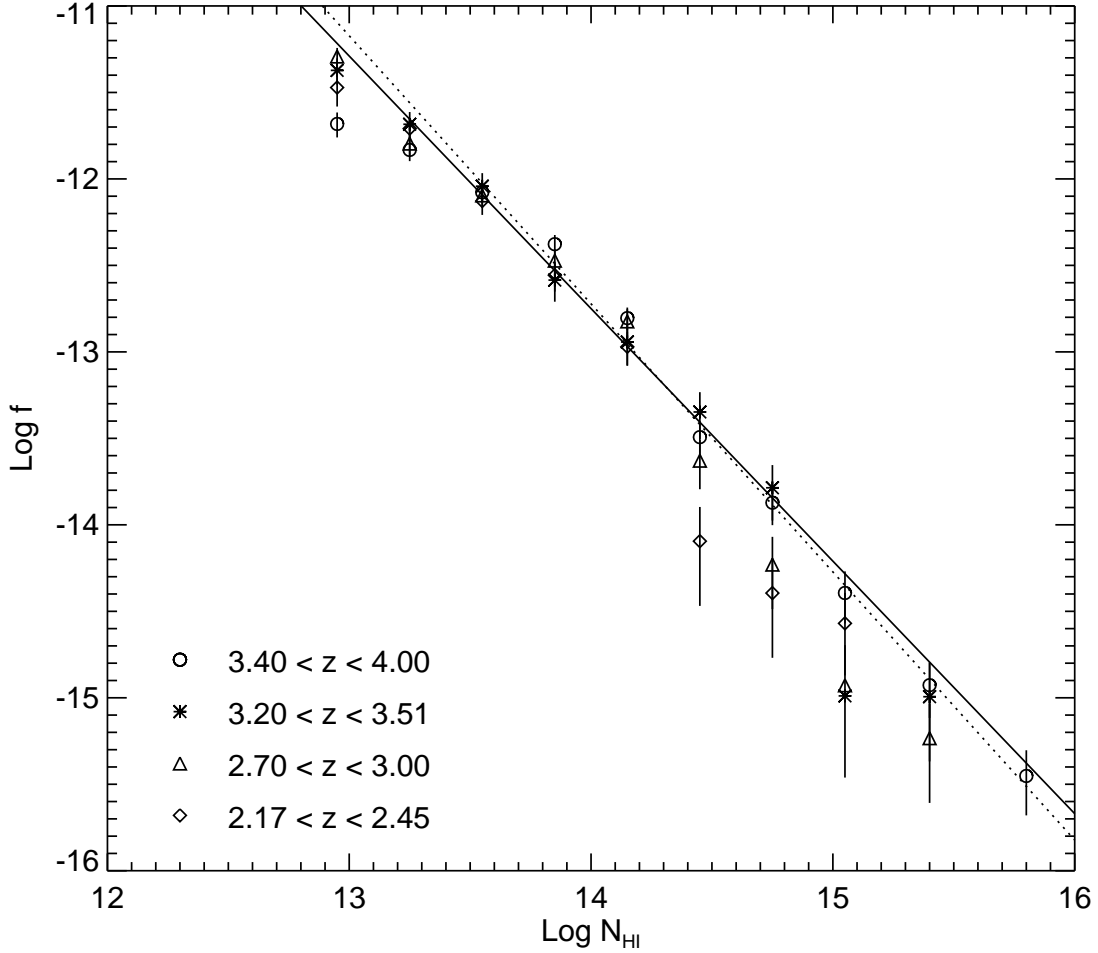}
\caption
{\label{fig:1}Differential density distribution function without
incompleteness correction. The distribution with \hi\ column density is
shown for each of three indicated redshift bins, with 1$\sigma$
error bars, compared with the results of \protect{\markcite{lu96}}Lu
\etal\ (1996) (open circles).  The solid line represents the single
power-law fit, $f(N_{\rm HI})=4.9 \times 10^{7}\ N_{\rm HI}^{-1.46}$.
The dotted line shows the single power-law fit taken from
\protect{\markcite{lu96}}Lu \etal\ (1996) for $<z> = 3.7$, which is
$f(N_{\rm HI}) \propto N_{\rm HI}^{-1.55}$ for $12.6 < \log N_{\rm HI} <
16.0$.  At $N_{\rm HI} \le 10^{14}\ {\rm cm}^{-2}$, the distribution
functions at different redshifts look very similar, and lack significant
flattening. However, at $N_{\rm HI} \ge 10^{14}\ {\rm cm}^{-2}$, the
amount of deficiency in forest clouds compared to the expected power-law
distribution depends on redshift. The deviation from the power law moves
to smaller $N_{\rm HI}$ as redshift decreases, which means the higher
$N_{\rm HI}$ forest clouds evolve rapidly.}
\end{figure}

Fig.~\ref{fig:1} shows the DDF, without any correction for blending or
incompleteness, at different redshifts. The stars represent $<z>=3.35$,
the triangles $<z>=2.85$, and the diamonds $<z>=2.31$.  The open circles
are for $<z>=3.70$ from \markcite{lu96}Lu \etal\ 1996.  The solid line
represents the single power-law fit, $f(N_{\rm HI})=4.9 \times 10^{7}
N_{\rm HI}^{-1.46}$, taken from \markcite{hu95}Hu \etal\ (1995) for their
data at $z \sim 2.85$.  The dotted line represents the single power-law
fit for $z \sim 3.7$ Ly$\alpha$ forest clouds, $f(N_{\rm HI}) \propto
N_{\rm HI}^{-1.55}$ estimated from Fig.~6 of \markcite{lu96}Lu \etal\ 
(1996), and matched to
the same $f(N_{\rm HI})$ at $\log N_{\rm HI} = 14.15$. This
slope of --1.55 at $<z>=3.7$ was obtained from the fit to the column
density range $N_{\rm HI} = 10^{12.6-16.0}\ {\rm cm}^{-2}$ with the
incompleteness correction applied.  However, over the column density
range $N_{\rm HI} = 10^{12.3-14.5} \ {\rm cm}^{-2}$, \markcite{lu96}Lu
\etal\ (1996) found a slope of --1.46, which matches the value found by
\markcite{hu95}Hu \etal\ (1995) and \markcite{kir97}Kirkman \& Tytler
(1997) at redshifts near 3.

There are two notable features of Fig.~\ref{fig:1}. First, the four DDFs
roughly follow a single power-law, $f(N_{\rm HI}) \propto N_{\rm
HI}^{-\beta}$, with $\beta \sim 1.4$ over the column density range
$N_{\rm HI} = 10^{12.8} - 10^{14.3}\ {\rm cm}^{-2}$. The DDFs do not show
any sign of flattening in this range with the possible exception of the
$<z>=3.70$ DDF at lower \hi\ column density. However, the lowest $N_{\rm
HI}$ points for \markcite{lu96}Lu \etal\ have higher levels of
incompleteness correction and this result is correspondingly uncertain.
Earlier results claimed that there was a flattening in the DDF at $N_{\rm
HI} \leq 10^{13.8}\ {\rm cm}^{-2}$ (\markcite{pet93}Petitjean
\etal\ 1993).  However, this flattening is mainly caused by blending in
low resolution data and was not confirmed from the Keck HIRES data down
to $N_{\rm HI} \sim 10^{12.5}\ {\rm cm}^{-2}$ at $z \sim 2.85$
(\markcite{hu95}Hu \etal\ 1995). We now see this result holds over the $z
= 2 - 4$ redshift range, and also provides a good description of the
\markcite{lu96}Lu \etal\ (1996) results.

Second, at the higher column density end, $N_{\rm HI} \ge 10^{14.3}\ {\rm
cm}^{-2}$, the DDFs for the redshift intervals differ substantially.
Except at the highest redshift there is a strong lack of forest clouds at
high column densities compared with the number expected from the
power-law distribution at $N_{\rm HI} \le 10^{14.3}\ {\rm cm}^{-2}$ (see
\markcite{pet93}Petitjean \etal\ 1993). This deficiency cannot be a
result of misestimates in the column density owing to saturation effects
since the total number of clouds is deficient relative to the
extrapolated power law (\markcite{hu95}Hu \etal\ 1995).  The $<z>=3.70$
data of \markcite{lu96}Lu \etal\ (1996) show only a slight steepening at
the higher column densities, but by $<z>=3.35$ forest clouds deviate from
the power-law at $N_{\rm HI} \sim 10^{14.8}\ {\rm cm}^{-2}$, while those
at $<z>=2.85$ and $<z>=2.31$ deviate from the power-law at $N_{\rm HI}
\sim 10^{14.3}\ {\rm cm}^{-2}$.  Furthermore, the degree of deficiency in
cloud numbers compared to predictions from the single power-law fit
also depends on redshift.  The $<z>=2.31$ clouds deviate from the
power-law more rapidly than the $<z>=2.85$ clouds and the $<z>=3.35$
clouds.  There are no clouds with $N_{\rm HI} \ge 10^{15.2}\ {\rm
cm}^{-2}$ at $<z>=2.31$.  The redshift path for $<z>=2.31$ is 1.5 times
larger than that for $<z>=3.35$, so this deficiency of higher column
density clouds in the lower redshift range cannot be an observational
selection effect.  Thus the break in the density distribution function
appears to be strengthening and migrating to lower column densities as
the redshift decreases.

\subsection{Number density of Ly$\alpha$ forest clouds with redshift}

The evolution in Ly$\alpha$ forest clouds with redshift has
traditionally been explored by looking at the number density of
clouds in a given column density range as a function of redshift.
The number density per unit redshift per line of sight is expressed
as $N(z) = N_{\rm o} (1+z)^{\gamma}$, where $N_{\rm o} = \Phi_{\rm
o}\pi R_{\rm o}^{2}cH_{\rm o}^{-1}$.  The parameter $\Phi_{\rm o}$ is
the local comoving number density of forest clouds, $\pi R_{\rm
o}^{2}$ is the local cross section of a cloud, and c and $H_{\rm o}$
are the speed of light and the Hubble constant, respectively.  This
yields $\gamma = 1$ for $q_{\rm o}=0$ and $\gamma = 0.5$ for $q_{\rm
o}=0.5$ for non-evolving clouds.  However, it is clear from the
evolution of the shape in the DDF that the measured $\gamma$ will be
a function of the chosen column density interval and of redshift,
except at low column density. This greatly complicates this type of
analysis, and comparisons between individual studies based on this
method are difficult.

HST observations show a value for $\gamma = 0.48 \pm 0.62$ at $z <
1$, which is consistent with no evolution (\markcite{bah93}Bahcall
\etal\ 1993).  However at higher redshift, $z > 2$, a rather wide
range of $\gamma$ values has been obtained, from $\gamma = 2.75 \pm
0.29$ (\markcite{lu91}Lu \etal\ 1991; indicating strong evolution) to
$\gamma = 2.36 \pm 0.40$ (\markcite{baj88}Bajtlik \etal\ 1988)
to $\gamma = 1.89 \pm 0.28$ (\markcite{bec94}Bechtold 1994;
indicating weak evolution). An inflection in $N(z)$ from the
redshift-limited samples has also been pointed out, suggesting that
$\gamma$ is different in different redshift ranges (\markcite{lu91}Lu
\etal\ 1991, \markcite{bec94}Bechtold 1994, \markcite{imp96}Impey
\etal\ 1996). These results have been used to suggest that Ly$\alpha$
forest clouds do not consist of a single population and that a
different population becomes dominant at different redshifts
(\markcite{bok95}Boksenberg 1995).

\begin{figure}
\plotone{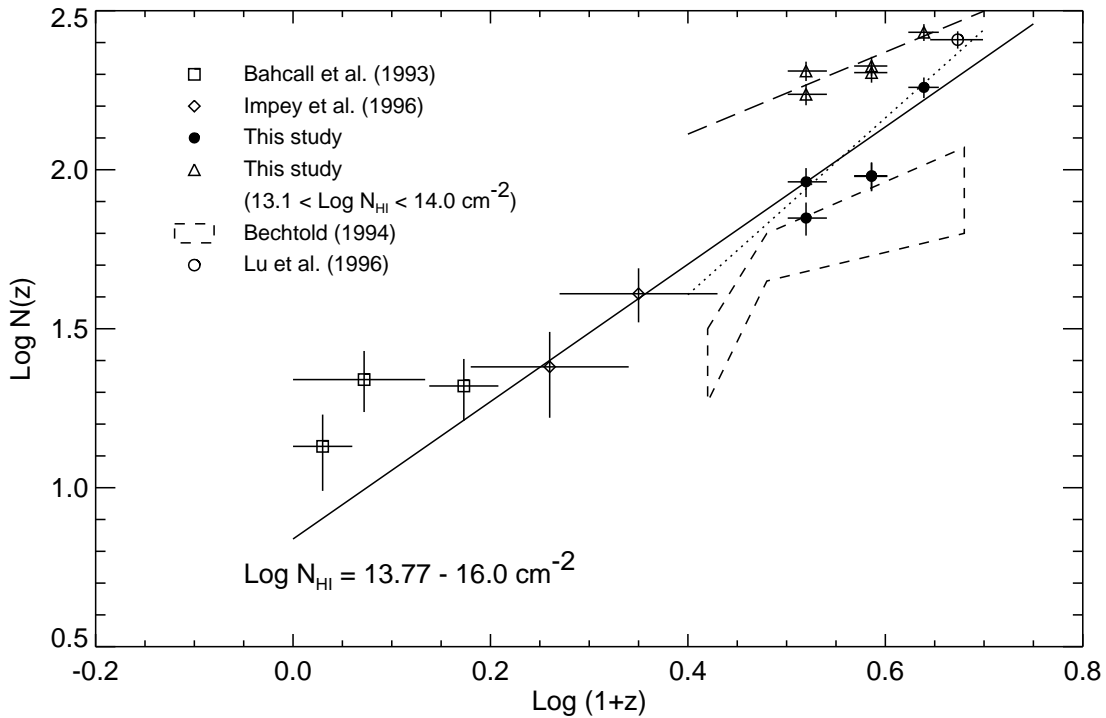}
\caption
{\label{fig:2}$N(z)$ vs redshift for the present data (filled
circles) and the HST observations (open squares and open diamonds).
The region outlined with dashed lines shows the results from
\protect{\markcite{bec94}}Bechtold (1994).  The solid line shows the
maximum likelihood fit to the HST data and the high column density
Keck data (slope $\gamma=2.15$).  The dotted line shows the fit to
the Keck data only, with a slope of $\gamma=2.78$.  Triangles show
the forest clouds over just the column density range $N_{\rm HI} =
10^{13.1}-10^{14}\ {\rm cm}^{-2}$, with a long dashed line fit with
$\gamma = 1.19$. The $1\sigma$ error bars and redshift binning are
indicated for each point.}
\end{figure}

Keeping in mind that $N(z)$ is sensitive to the adopted threshold
$N_{\rm HI}$ (a reflection of the form of departure from the
single-power law distribution at $N_{\rm HI} \ge 10^{14.3}\ {\rm
cm}^{-2}$), we fitted the data to a conventional single power-law
form, for lines with $N_{\rm HI} = 10^{13.77} - 10^{16}\ {\rm
cm}^{-2}$. This corresponds roughly to the conventional equivalent
width threshold, $W=0.32 {\rm \AA}$, and enables us to compare the
present data directly with the results of \markcite{bec94}Bechtold
(1994) and \markcite{imp96}Impey \etal\ (1996).  Fig.~\ref{fig:2}
shows $N(z)$ as a function of redshift.  The vertical bars indicate
the 1$\sigma$ errors and the horizontal bars indicate the bin sizes
for the individual QSOs.  No incompleteness correction was made; over
this column density range, it is rather small (less than 10\%) and
within the 1$\sigma$ error bars.  The filled circles show the Keck
HIRES data for our 5 QSOs and the open circle that of \markcite{lu96}Lu
\etal\ (1996).  The open triangles show the Keck HIRES
data only for the lower column density systems with $N_{\rm HI} =
10^{13.1} - 10^{14.0}\ {\rm cm}^{-2}$.  The diamonds are taken from
\markcite{imp96}Impey \etal\ (1996) for HST observations of
PG1222+228 and PG1634+706 and the open squares are taken from
\markcite{bah93}Bahcall \etal\ (1993) for HST observations of PKS
0044+03, 3C 273, and 3C 351.  The region shown by dashed lines
represents the measurement of \markcite{bec94}Bechtold (1994), which
is a factor of 2 lower than our observations. This is probably
explained by the 7 times higher resolution in our data, which is
capable of deblending neighboring lines which, especially at higher
redshifts, can be a serious problem.  Our data yield a factor of 2.3
more lines at $z \sim 3.2$ and a factor of 1.3 more lines at $z \sim
2.4$ than found by \markcite{bec94}Bechtold (1994), consistent with
this interpretation of the discrepancy.

The maximum likelihood method (\markcite{mur86}Murdoch \etal\ 1986) was
used to derive the best estimated $\gamma$ values for the above data.
The best fit using only the present Keck data in our study at $3.5 > z >
2$ (the dotted line) yields $\gamma = 2.78 \pm 0.71$, consistent with
$\gamma = 1.89 \pm 0.28$ by \markcite{bec94}Bechtold (1994) within a
2$\sigma$ error.  The best fit for $N_{\rm o}$ is $N_{\rm o} = 3.12$.
The uncertainty in $N_{\rm o}$ was not calculated since it is strongly
correlated with $\gamma$.  At lower redshifts, $z<1.5$, line blending
becomes less of a problem than for $z>2$ systems, since the number
density of clouds is smaller, and the HST counts should be directly
relatable to the Keck counts. When our data is combined with the HST
observations (\markcite{bah93}Bahcall \etal\ 1993, \markcite{imp96}Impey
\etal\ 1996), the best fit for $0<z<3.6$ yields $\gamma = 2.15 \pm 0.21$
and $N_{\rm o} = 6.89$ (the solid line).  Because there are only 15
forest clouds at $z < 1$, the maximum likelihood method puts high
emphasis on the numerous and more densely covered (in redshift space)
high-redshift observations, resulting in a higher $\gamma$ than that
obtained with a least-squares fit, $\gamma = 1.74 \pm 0.14$.  The
$\gamma$ value driven by the maximum likelihood method is also higher
than the one ($\gamma = 0.48$) estimated from the $z<1$ data
(\markcite{bah93}Bahcall \etal\ 1993) or the one ($\gamma = 1.26$) from
the $0<z<4$ data obtained by \markcite{bec94}Bechtold (1994).  The latter
result is a consequence of Bechtold's lower number counts.  The value of
$N_{\rm o}$ is also correspondingly smaller (by a factor of 3) than that
obtained by \markcite{bah93}Bahcall \etal\ (1993) ($N_{\rm o} \sim
19.0$) because of the higher $\gamma$. The present high $\gamma$ values show
that the number of higher column density clouds at $z>1$ declines rapidly
with decreasing redshift.  When the $<z>=3.7$ systems are included, the
value of $\gamma$ for $0<z<4$ is $2.41\pm0.18$.  However, considering the
different treatment applied to the different data sets, this estimated
value should be taken with caution.

When we count only Ly$\alpha$ forest clouds in the column density
range of $N_{\rm HI} = 10^{13.1} - 10^{14}\ {\rm cm}^{-2}$, where the
DDF is well-fitted by a common power-law for systems at all
redshifts, we obtain a best fit (the long-dashed line) $\gamma = 1.29
\pm 0.45$ and $N_{\rm o} = 39.44$ for the $z > 2$ systems.  This
$\gamma$ value is consistent within the errors with a non-evolving
model ($\gamma= 1$), as we would expect from our discussion of
the DDF.  Although the lack of lower column density forest clouds at
$z<1$ prevents extrapolation of these fitted values, it appears that
the DDF and $N(z)$ of these lower column density clouds do not evolve 
with redshift at $z>2$.

\subsection{The change in {$b$} value with redshift\label{subsec:bvalue}}

\begin{figure}
\figurenum{3}
\label{fig:3}
\plotone{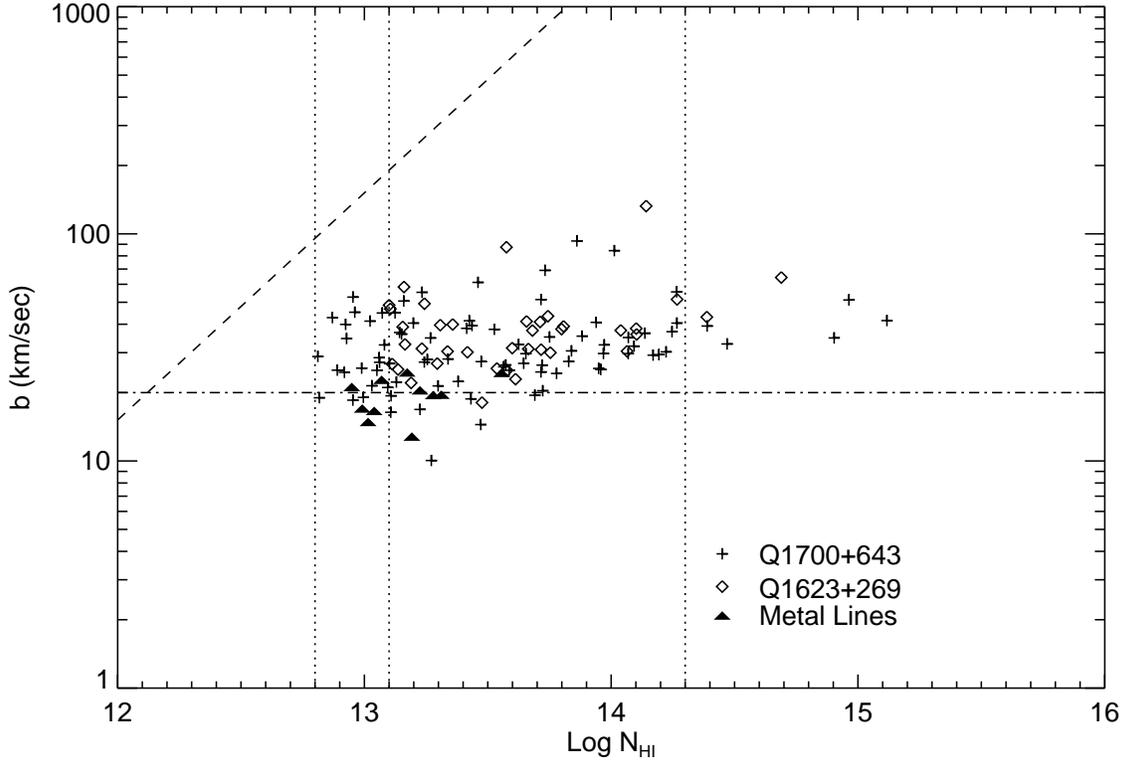}
\figurenum{3a}
\caption
{$b - \log N_{\rm HI}$ diagram at $<z>=2.17 - 2.45$.
The dot-dashed line shows $b = 20~{\rm km\ s}^{-1}$. Note that
there are not many lines with $N_{\rm HI} \ge 10^{14}\ {\rm cm}^{-2}$.
The vertical dotted lines show the adopted minimum $N_{\rm HI}$ and the value
at which the clouds saturate, while the dashed line shows the optical
depth selection limit.}
\end{figure}

The line widths are characterized by the Doppler parameter, $b$, in the
profile fitting.  Figs.~\ref{fig:3}a--c show $b - \log N_{\rm HI}$
diagrams at average redshifts $<z>=2.31$, $<z> =2.85$, and $<z>=3.35$.  As
is shown in the diagrams, most lines with $b < 15~{\rm km\ s}^{-1}$ are
metal lines.  There seems to be a reasonably well-defined minimum $b$
value ($b_{\rm c}$) over the $N_{\rm HI}$ range at each redshift when the
metal lines are discarded (see also \markcite{hu95}Hu \etal\ 1995 and
\markcite{lu96}Lu \etal\ 1996, though this result is questioned by
\markcite{kir97}Kirkman \& Tytler 1997),
implying that the $b$ value distribution is
truncated at the low end.  (Note the increase in the number of lines with
$N_{\rm HI} \ge 10^{14}\ {\rm cm}^{-2}$ with increasing redshift.) Also
the number of clouds with $b \le 20~{\rm km\ s}^{-1}$ (the $b_{\rm c}$
value at $z \sim 2.85$ shown as a dashed line) decreases as the redshift
decreases, suggesting that $b_{\rm c}$ is increasing with decreasing
redshift (see also \markcite{lu96}Lu \etal\ 1996).

In order to quantify the changes in $b_{\rm c}$ and determine the
corrections to the DDF for line blending and incompleteness and the
profile fitting methodology, we followed the procedures of
\markcite{hu95}Hu \etal\ (1995). We generated artificial spectra and
fitted them in the same way as we treat the observed data. We assumed
the slope in the DDF to be 1.5 in order to compare to the $<z>=2.85$
clouds (\markcite{hu95}Hu \etal\ 1995), and used a random
distribution of clouds at a given redshift range. For $b$
values, we used a Gaussian distribution, but truncated it at a value
lower than the cut-off value, $b_{\rm c}$, at each redshift.  The fitted
artificial counts are then compared to the input counts to determine the
changes caused by blending. This incompleteness correction at $<z> =
2.31$ and at $<z> = 3.35$ as a function of column density is given in
Table \ref{tbl-2}, while the incompleteness correction at $<z>=2.85$ is 
taken from \markcite{hu95}Hu \etal\ (1995).  Fig.~\ref{fig:4} shows the
incompleteness corrected differential density distribution functions.
Since the line density is larger at higher redshift, the incompleteness
correction at $<z>=3.35$ is slightly more sensitive to the assumed
$\beta$ and the fitted $N_{\rm HI}$ range.  Assuming $\beta = 1.5$, the
resulting $\beta$ is $\sim 1.58$ for $N_{\rm HI} =
10^{12.8}-10^{14.3}\ {\rm cm}^{-2}$ and $\sim 1.53$ for $N_{\rm HI} =
10^{12.8}-10^{14.8}\ {\rm cm}^{-2}$.  The incompleteness correction at
$<z>=2.31$ is not as sensitive to the assumed $\beta$ as the $<z>=3.35$
clouds. Here, the observed low $\beta$ of 1.28 results in a low $\beta$
of 1.35 after the incompleteness correction.  For $<z>=3.35$, $f(N_{\rm
HI})=3.3 \times 10^{9} N_{\rm HI}^{-1.59 \pm 0.13}$ (2$\sigma$ error).
For $<z>=2.31$, $f(N_{\rm HI})=1.5 \times 10^{6} N_{\rm HI}^{-1.35 \pm
0.03}$ (2$\sigma$ error).  We note that using only the two $<z>$=2.85 systems,
Q0014+813 and Q0302--003, gives an incompleteness corrected DDF having
a best fit of $f(N_{\rm HI})=5.7 \times 10^{6} N_{\rm HI}^{-1.39 \pm0.26}$
(2$\sigma$ error) over the column density range $N_{\rm HI} = 
10^{12.8}-10^{14.3}\ {\rm cm}^{-2}$.  This is slightly shallower
than the $-1.46$ slope obtained by \markcite{hu95}Hu \etal\ (1995) with
two additional quasars included for this redshift interval. 
Although there might be some deviation at the
lower $N_{\rm HI}$ end in the distribution function (where the
incompleteness correction is very uncertain), the three DDFs show an
excellent match at $N_{\rm HI} = 10^{13.7}-10^{14.3}\ {\rm cm}^{-2}$
where the incompleteness correction is not high. Within the 2$\sigma$
errors and the systematic errors in the profile fitting methodology,
these $\beta$ values could be treated as consistent with the $<z>=2.85$
forest clouds with $\beta = 1.46 \pm 0.07$, but overall there does seem
to be evidence that there is a very weak flattening of $\beta$ with
decreasing redshift. However, we conclude again that unlike the case for
high column density clouds ($N_{\rm HI} \ge 10^{14.3}\ {\rm cm}^{-2}$),
the DDF does not change rapidly with redshift, within the observational
and simulated errors for $N_{\rm HI} \le 10^{14.3} \ {\rm cm}^{-2}$.

The procedure of incompleteness correction also produces an estimated
$b_{\rm c}$ value. With an assumed $\beta$, three parameters of a
truncated Gaussian $b$ distribution are varied until a satisfactory
match to the observations is obtained:  median($b$), 1$\sigma(b)$,
and $b_{\rm c}$.  Fig.~\ref{fig:5} shows the averaged simulated $b$
distribution (the solid lines) with the observed $b$ distribution
(the dotted line) at $<z>=2.31$ and $<z>=3.35$. With the large number
of lines at $<z>=3.35$, the cut-off $b$ value, $b_{\rm c}$, is rather
straightforward to determine: $b_{\rm c} = 17~{\rm km\ s}^{-1}$ (or
$T_{\rm c} = 1.8 \times 10^{4}$ K, if we assume a thermal
broadening), 1$\sigma = 8~{\rm km\ s}^{-1}$, and median($b$) =
$26~{\rm km\ s}^{-1}$. No perfect fit was found for the $<z>=2.31$
forest clouds. With $b_{\rm c} = 22~{\rm km\ s}^{-1}$, more than 50\%
of the simulations generated too many lines with $b = 19-22~{\rm
km\ s}^{-1}$ when compared to the observations. Assuming that most
narrow lines are metal lines (90\% of the unidentified lines with $b
\le 20~{\rm km\ s}^{-1}$ are from the Q1700+643 with 10 metal line
systems), we took the best fit value for the observation namely at $b
\ge 20~{\rm km\ s}^{-1}$: $b_{\rm c} = 24~{\rm km\ s}^{-1}$ (or
$T_{\rm c} = 3.5 \times 10^{4}$ K, if thermally broadened), 1$\sigma
= 12~{\rm km\ s}^{-1}$, and median($b$) = $30~{\rm km\ s}^{-1}$.

These results show that there is an increase in the $b_{\rm c}$ value
as redshift decreases which is also consistent with $b_{\rm c} =
15~{\rm km\ s}^{-1}$ determined at $z = 3.4 \to 4.0$ by
\markcite{lu91}Lu \etal\ (1996).  These results are shown in
Fig.~\ref{fig:6a}. Fig.~\ref{fig:6b} shows the median $b$ values for
each QSO. The solid line is for $N_{\rm HI} = 10^{13.8} - 10^{16}
\ {\rm cm}^{-2}$ and the dotted line is for $N_{\rm HI} = 10^{13.1} -
10^{14} \ {\rm cm}^{-2}$. Table \ref{tbl-3} lists the median $b$ values for
each QSO.  This rising trend in both $b_{\rm c}$ and the median $b$
values suggests that the internal temperatures and/or the level
of kinematic broadening are rising with decreasing redshift, as we shall
discuss further in Sec.\ \ref{sec:discuss}.

\subsection{Correlation Function}

The simplest way of testing inhomogeneity in the space distribution
of Ly$\alpha$ forest clouds is to calculate the two-point velocity
correlation function, $\xi$. Among different ways of calculating
$\xi$, the most widely used method is comparison between the observed
number of pairs ($N_{obs}$) and the expected number of pairs
($N_{exp}$) from the random distribution in a given velocity bin
($\Delta v$); $\xi(\Delta v) = N_{obs}(\Delta v)/N_{exp}(\Delta v)
-1$, where $\Delta v = c(z_{2}-z_{1})/[1+ (z_{2}+z_{1})/2]$ and
$z_{1}$ and $z_{2}$ are the redshifts of two lines.  Although this
formula is correct only in the limit of an ideal large sample and is
sensitive to the bin size, general information on clustering can be
obtained by this method.

Searches for clustering in Ly$\alpha$ forest clouds have been 
a function of resolution, S/N, bin size, and possibly line of sight,
and there have been rather mixed results to date.
\markcite{sar80}Sargent \etal\ (1980) and \markcite{rau92}Rauch
\etal\ (1992) find no positive correlation at any velocity scale. On
the other hand, \markcite{cri95}Cristiani \etal\ (1995), and
\markcite{hu95}Hu \etal\ (1995) show an anti-correlation at $\Delta v
= 600 - 900~{\rm km\ s}^{-1}$ as well as a positive correlation at
$\Delta v < 200~{\rm km\ s}^{-1}$. While this anti-correlation is
also claimed to exist by \markcite{mei95}Meiksin \& Bouchet (1995),
\markcite{cri97}Cristiani \etal\ (1997) failed to confirm the reality
of anti-correlation from a larger data set, but did confirm the
positive correlation at $\Delta v < 300~{\rm km\ s}^{-1}$.

The correlation functions at the three redshifts were calculated
relative to the average of 500 simulations with a $50~{\rm
km\ s}^{-1}$ velocity bin for each QSO.  We used lines with $N_{\rm
HI} \le 10^{16}\ {\rm cm}^{-2}$ for this calculation, although there
are few lines with $N_{\rm HI} \ge 10^{15} \ {\rm cm}^{-2}$.

Fig.~\ref{fig:7a} shows the averaged correlation function as a
function of redshift for systems with column densities above
a threshold value, $N_{\rm HI, th}$, such that
$\log N_{\rm HI, th} > 13.8$. The dotted
lines are the 1$\sigma$ Poisson errors. While there is suggestion of
a positive correlation for $<z>=2.31$ at $\Delta v < 300~{\rm
km\ s}^{-1}$ (a 2.8$\sigma$ significance at $\Delta v = 75~{\rm
km\ s}^{-1}$), there is no significant correlation strength at $<z> =
2.85$. This is in disagreement with what was found by
\markcite{hu95}Hu \etal\ (1995) with an overlapping HIRES data set.
We will discuss this point later.  At $<z>=3.35$, a positive
correlation may exist at a 2.54$\sigma$ significance at $\Delta v =
75~{\rm km\ s}^{-1}$. Given the large time gap between $z = 3.35$ and
the present day, the interpretation of the correlation functions with
redshift is ambiguous. However, the clustering should evolve strongly
with redshift at $z = 2.31 \to z = 0$ if Ly$\alpha$ forest clouds
follow more or less the highly structured present day galaxies
(\markcite{her96}Hernquist \etal\ 1996, but see
\markcite{mor93}Morris \etal\ 1993).  Fig.~\ref{fig:7b} shows the
correlation strength at $\Delta v = 75~{\rm km\ s}^{-1}$ as a
function of threshold column density $N_{\rm HI,th}$, for three redshift bins. 
As \markcite{cri97}Cristiani \etal\ (1997) pointed out, at $<z>=2.31$
and $<z>=3.35$, the correlation strengths seem to increase as $N_{\rm
HI,th}$ increases. At $<z>=2.85$, this trend does not hold tightly.
However, as Fig.~\ref{fig:7c} shows, when averaged for all redshifts,
the correlation strength does increase as $N_{\rm HI,th}$ increases.

\subsection{Alternative Descriptions\label{sec:alt}}

Figs.\ \ref{fig:8a} and \ref{fig:8b} show two alternative descriptions of
the data (\eg, \markcite{cro97}Croft \etal\ 1997).  Fig.\ \ref{fig:8a}
shows the normalized cumulative distribution of the flux decrements $D
\equiv 1 - e^{\tau}$ for the five quasars computed in the wavelength ranges
specified in Table \ref{tbl-1}.  The distributions agree well with similar
observations by \markcite{rau96}Rauch \etal\ (1996) and also with model
predictions for a suitable choice of $\Omega_{\rm b}\,h^2\,T^{-0.7}/J$
(\eg, \markcite{rau96}Rauch \etal\ 1996, \markcite{cro97}Croft
\etal\ 1997).  Fig.\ \ref{fig:8b} shows the number of times the specrum
crosses down across a given decrement per unit $X$.  In the absence of
blending and given the invariance of the $DDF$, narrower $b$ values will
result in a higher number of crossings.  Thus, the increased rate of
crossings seen at higher $z$ for the larger flux decrements is an
alternative method of seeing that the lower $z$ clouds are broader.  At
small flux decrements blending becomes important, and because this effect
is larger at higher $z$, this reverses the sign of the effect.

\section{Discussion\label{sec:discuss}}

\subsection{The column density distribution function}

The number density per unit redshift path $X$ per unit $N_{\rm HI}$ (the
density distribution function) as a function of redshift depends on
the history of the UV ionizing background, the mean baryon density,
and the underlying cosmology, and only detailed numerical
simulations can give complete descriptions of these combined effects to
compare with the observations.

Simulations by \markcite{muc96}M\"{u}cket et al.\ (1996) show broad
consistency with the present data, with high column density clouds
($N_{\rm HI} > 10^{14}\ {\rm cm}^{-2}$) evolving as $16.8(1+z)^{0.8}
+ 0.687(1+z)^{2.5}$ while low column density clouds ($N_{\rm HI} <
10^{13}\ {\rm cm}^{-2}$) evolve only as $(1+z)^{0.7}$.  However, 
Fig.\ \ref{fig:9} shows that the observed rise for the high column
density ($10^{14} < N_{\rm HI} < 10^{16} \ {\rm cm}^{-2}$) clouds is
somewhat steeper than this prediction.  \markcite{lu96}Lu \etal's
results at $<z>=3.7$ are also shown, and are consistent with the
fit obtained for $<z>=2.31 \to 3.35$ .  The more sophisticated treatment of
\markcite{dav97}Dav\'e \etal\ (1997), using Voigt profile fitting to
artificial spectra generated from their SPH simulations, shows a DDF
that drops by a factor of $1.5 - 2$ for $z= 3 \to 2$ at $N_{\rm HI} \ge
10^{13.2}\ {\rm cm}^{-2}$, while at $N_{\rm HI} \le 10^{13.2}\ {\rm
cm}^{-2}$, the number of clouds increases by a factor of 2 for $z=3 \to
2$. These features in the predicted DDF do not match the observations,
which show a larger number of clouds at $z \sim 3$ compared to those at
$z \sim 2$ only for those clouds with $N_{\rm HI} \ge 10^{14.3}\ {\rm
cm}^{-2}$, but this may be due at least in part to the profile fitting
methodology.  Both \markcite{dav97}Dav\'e \etal\ (1997) and
\markcite{wei97}Weinberg \etal\ (1997) find relatively steep profiles
with $\beta=1.7$ as do \markcite{muc96}M\"{u}cket et al.\ (1996).
However, \markcite{zha97}Zhang \etal's (1997) similar analysis of their
hierarchical grid code simulation shows a remarkably good agreement
both with the overall shape of the DDF at $z=2.9$ ($\beta\approx1.5$),
and with the redshift evolution seen in the present data, finding
$\gamma$ for $W \ge 0.32 {\rm \AA}$ selected clouds of $2.95 \pm 0.4$
for $z=1.5\to3.5$ and $0.24 \pm0.03$ for $z<1.5$, which is in good
agreement with Fig.\ \ref{fig:2} and the discussion of
Sec.\ \ref{subsec:bvalue}.  It is unclear whether the differences
between \markcite{dav97}Dav\'e \etal\ and \markcite{zha97}Zhang
\etal\ (which are also reflected in the $b$ value distributions
discussed below) are a consequence of details of the codes, choice of
parameters (particularly the fluctuation spectra), or details of the
profile fitting analysis, but the results do suggest that within the
tunability and uncertainty in the current models that it is possible to
obtain remarkably good agreement with the data.

The analytic models of \markcite{hui96}Hui \etal\ (1996) give considerable
insight into the origin and invariance of the slope.  At these low column
densities the temperature is set by the balance between adiabatic
expansion and ionization heating, and may be described by a polytropic
equation of state $T \sim \rho^{\gamma -1}$, where $(\gamma-1)$ lies
between 0 and 0.62, depending on the ionization history.  The slope of the
DDF is then given by
\begin{equation}
  \beta = 1 + {\left [ \displaystyle { 0.96 - 2 \left ( \sigma_{\rm o} - 1 
         \right ) } \over \displaystyle {1.68 - 0.7  (\gamma -1)} \right ] }
\end{equation}
where $\sigma_{\rm o}$ is the RMS density fluctuation at scales close to the
Jeans length (\markcite{hui96}Hui \etal\ 1996).  For the observed $\beta$ 
values and for the possible range of $\gamma$ values, $\sigma_{\rm o}$
is extremely tightly constrained to a value of 1.15, with a plausible
range of only $\pm0.1$.  The flattening of the $\beta$ value with
decreasing redshift would require $\gamma$ to be closer to isothermal
at the lower redshifts, which is counter to the expected evolutionary
trend.  This may, however, be related to reionization of the He if this
occurs around $z=3$ as has been suggested based on the evolution of
metal line ratios (\markcite{son96}Songaila \& Cowie 1996).

The deviation from the power law at the high column density end would
then correspond to the overdensity at which shocking occurs (\eg,
\markcite{muc97}M\"{u}cket \etal\ 1997), which is approximately ten
times the mean density at the given redshift.  If the column density at
which the deviation occurs does correspond to the shocking overdensity
then we expect it to scale roughly as $(1+z)^6\,T^{-0.7}J^{-1}$, where
$T$ is the temperature and $J$ the ionizing flux.  If the ionizing flux
is roughly constant over the redshift range (\markcite{har96}Haardt \&
Madau 1996) and the temperature changes only slowly with redshift as in
the limit of early ionization (\markcite{hui96m}Hui \& Gnedin 1996),
$N_{\rm dev}$ would drop by 0.8 in the log between $z=3.5$ and $z=2.3$,
which is somewhat larger than the observed drop from $\log\,N_{\rm
HI}=14.8$ to 14.3.  This suggests that $J$ is falling with decreasing
redshift or that the temperature is rising.  The latter effect could
again be associated with helium reionization.

\subsection{The evolution of {\it b} values\label{subsec:bevol}}

If the Ly$\alpha$ forest lines were intergalactic clouds in thermal
equilibrium with the metagalactic UV ionizing background without any
other energy source, the line width of the clouds (or the Doppler
parameter, $b$) would measure the temperature of the clouds.  However,
within the structure evolution models the low column density forest
clouds ($N_{\rm HI} < 10^{14}\ {\rm cm}^{-2}$) are primarily
kinematically rather than thermally broadened.  Two important
observations may be taken from the Keck HIRES data. First, there seems
to be a rough lower cut-off $b$ value ($b_{\rm c}$) which increases
with decreasing redshift.  Most forest clouds have a $b$ value similar
to or greater than $b_{\rm c}$, whereas most lines with $b < b_{\rm c}$
are identified as metal lines.  \markcite{hu95}Hu \etal\ (1995)
interpreted $b_{\rm c}$ as the minimum $b$ value in a single region,
with most Ly$\alpha$ clouds being a blend of several components.
Secondly, taking the median $b$ as $34~{\rm km\ s}^{-1}$ and $b_{\rm
c}=20~{\rm km\ s}^{-1}$ suggests that the additional broadening due to
the multiple components is very similar to $b_{\rm c}$
(\markcite{cow95}Cowie \etal\ 1995).  The increase in the $b$ values
with decreasing redshift is likely to be a reflection of the ongoing
increase in structure.

The numerical simulations of \markcite{dav97}Dav\'e \etal\ (1997) give higher
$b$ values than are observed at $z\sim3$, while the simulations of
\markcite{zha97}Zhang \etal\ (1997) provide fairly good agreement with the
observations.  (Both are broader than the distribution of
\markcite{kir97}Kirkman \& Tytler 1997.)\ \ This is a similar effect to
that seen in the DDF, and may indicate some fundamental difference
between the SPH and grid simulations.  There is as yet no discussion of
the evolution of the $b$ value distribution, though the overall increase
in the width with redshift is in the sense expected from the models.

For the higher column density clouds which have entered the shocked
regime the widths will reflect primarily the thermal broadening, and the
minimum widths may then correspond to minimum temperatures (\eg,
\markcite{hae96}Haehnelt \etal\ 1996).  However, even for these clouds
the evolution of internal temperatures governed solely by ionization
equilibrium is insufficient to provide an explanation of the evolution of
$b_{\rm c}$.  \markcite{son96}Songaila \& Cowie (1996) show that
\siliv/\civ\ in the forest clouds at $2.1 <z< 3.8$ is well described by a
hypothetical \heii\ ionization front at $z \sim 3.1$, with Si/C = 3 times
the solar value, since the ratio of \siliv/\civ\ is sensitive at the
high energy end in the assumed metagalactic UV flux. If there is a
\heii\ ionization front at $z \sim 3.1$, this ionization causes a change
in the metagalactic UV flux, and thus the $b$ value in the forest clouds.
We tested this effect using a broken power law for the UV flux, following
the same procedure as \markcite{gia94}Giallongo \& Petitjean (1994),
where the amount of \heii\ ionization is characterized by a parameter B,
which is the relative flux at 4 ryd and 4.5 ryd. Assuming the flux at the
Lyman Limit (1 ryd) of $J_{\rm o} = 10^{-21}$ erg cm$^{-2}$ s$^{-1}$
Hz$^{-1}$ and a single power-law for the UV energy spectra with $\alpha =
1.5$ ($J_{\nu} \propto \nu^{-\alpha}$) (\markcite{coo97}Cooke
\etal\ 1997), we ran the photoionization model of \markcite{fer93}Ferland
(1993). The calculation was made for clouds with $N_{\rm HI} =
10^{14.5}\ {\rm cm}^{-2}$ and we used a $10^{-2}$ solar metallicity with
Si/C = 3. Fig.~\ref{fig:10} shows the model calculations with different
ionization parameters, $\Gamma \sim J_{\rm o} / n_{\rm e}$.  The
long-dashed line is for $B=1$ (no \heii\ ionization); the dotted line,
the dashed line, the dot-dashed line, and the dot-dot-dashed line are for
$B = 10$, $B=100$, $B= 1000$, and $B=10000$, respectively.
\markcite{son96}Songaila \& Cowie (1996) found a satisfactory result for
the Si and C ratios with $B \ge 100$ for $z > 3.1$ and $B = 1 - 10$ for
$z < 3.1$.  At $<z>= 2.85$, the model-calculated $b$ for the hydrogen
clouds is well-matched with the $B = 1-10$ spectra at $\Gamma = 10^{-2} -
10^{-1.5}$ and at at $<z>=3.35$ and 3.7, the $b_{\rm c}$ values are also
in the regime of the photoionization prediction with $B = 100$.  However,
at $<z>=2.31$, the $b_{\rm c}$ value is too high for the photoionization
model to obtain (for either $b_{\rm c} = 22~{\rm km\ s}^{-1}$ or $b_{\rm
c} = 24~{\rm km\ s}^{-1}$).  Thus evolution in $b_{\rm c}$ cannot be
fully accounted for by evolution of an equilibrium photoionization
temperature.  A hypothetical \heii\ ionization at $z \sim 3.1$ can
explain the evolution of $b_{\rm c}$ values at $z > 2.5$, but fails to
explain the $b_{\rm c}$ of $24~{\rm km\ s}^{-1}$ at $<z>=2.31$.   The
lower redshift clouds clearly require a higher dynamical energy input
than the high redshift clouds.  This may reflect a higher level of overall
turbulence at the lower redshifts, as evidenced by the increase in the
median $b$ values.

\subsection{Ly$\alpha$ forest contribution to $\Omega_{\rm b}$}

One of the most important cosmological parameters is the ratio of
the baryon density to the critical density in the universe, $\Omega_{\rm b}$,
which can be directly compared with the predictions of big bang
nucleosynthesis.  At $z= 2-4$ the intergalactic gas may still comprise
most of the baryons in the universe and thus the QSO absorption line
systems can provide a direct measurement of $\Omega_{\rm b}$.

Numerical simulations of the Lyman forest have generally favored a high
$\Omega_{\rm b}h^2$ ($\approxgt0.0125$) (\eg, \markcite{rau96}Rauch
\etal\ 1996, \markcite{wei97}Weinberg \etal\ 1997), though a more direct
study of the \heii\ opacity suggests a lower $\Omega_{\rm b}$
(\markcite{hog97}Hogan \etal\ 1997).  All of these methods are dependent
on assumptions about the evolution of the shape and normalization of the
ionizing field (as well as its fluctuations), which determine the cloud
internal temperatures and the conversion from neutral to total hydrogen.
It is also possible that the current simulations may not be adequately
dealing with the structure on the smaller scales (\markcite{cro97}Croft
\etal\ 1997), which again could result in uncertainties in the relative
normalization of the hydrogen optical depth relative to $\Omega_{\rm
b}$.  Here we shall adopt a more empirical approach to determining the
evolution of $\Omega_{\rm b}$ from the observations.

The contribution to $\Omega_{\rm b}$(\hi) from the Ly$\alpha$ forest
can be directly obtained by integrating through the DDF.  In units of
the local critical density $\rho_{\rm c}$, 
the mass density of \hi\ in the Ly$\alpha$ forest is
\begin{equation}
 \Omega_{\rm b}({\rm HI}) = {H_{\rm o}\over c} {\mu m_{\rm H} \over
  \rho_{\rm c}} \int_{N_{\rm min}}^{N_{\rm max}}
  N_{\rm HI} f(N_{\rm HI}) dN_{\rm HI}, 
\end{equation}
(\eg, \markcite{lan91}Lanzetta \etal\ 1991).
The parameter $m_{\rm H}$ is the hydrogen mass, c is the speed of
light, $H_{\rm o}$ is the Hubble constant, and $N_{\rm min}-N_{\rm max}$ is the
\hi\ range of the forest clouds. 
The parameter $\mu$ is the mean mass per proton mass and is taken as 1.3.

Because the DDF is everywhere shallower than a --2 power-law slope,
$\Omega_{\rm b}$(\hi) is always dominated by the highest column density
clouds.  In Table \ref{tbl-4} we summarize $\Omega_{\rm b}$(\hi) from clouds 
below $2 \times 10^{14}\ {\rm cm}^{-2}$, which is always below the break, and
also the total $\Omega_{\rm b}$(\hi) for $N_{\rm HI} < 4 \times
10^{15}\ {\rm cm}^{-2}$.  In the lower $N_{\rm HI}$ range which is below
the break, $\Omega_{\rm b}$(\hi) is roughly constant at $6 \times
10^{-8}\,h^{-1}$ over the redshift range, reflecting the rough invariance
of the DDF, while in the higher column densities $\Omega_{\rm b}$(\hi)
drops from $2.5 \times 10^{-7}\,h^{-1}$ at $z=3.7$ to $1.1 \times
10^{-7}\,h^{-1}$ at $z=2.31$, reflecting the decreasing number of clouds.

In order to translate $\Omega_{\rm b}$(\hi) to $\Omega_{\rm b}$ we need to
know the appropriately weighted ionization fraction $F=$ \hi/H in the
clouds.  At the high column density end ($N_{\rm HI} \approxgt 10^{15}\ {\rm
cm}^{-2}$) the ionization state of the clouds is directly probed by
metallic line ratios and \cii/\civ\ ratios, around 0.01
(\markcite{son96}Songaila \& Cowie 1996), give typical values of the
ionization parameter $U=0.002\to 0.03$, where $U = 4.2\times
10^{-5}\ J_{21}/n$ in terms of $n$, the total hydrogen density, and
$J_{21}$, the ionizing flux in units of $10^{-21}$ ergs cm$^{-2}$ s$^{-1}$
Hz$^{-1}$ sr$^{-1}$. Here we have assumed $J_{\nu} \sim \nu^{-\alpha}$ with
$\alpha=+1.5$, following the discussion of \markcite{don91}Donahue \& Shull
(1991).  The exact $U$ value determined from the \cii/\civ\ ratio depends
on the details of the cloud temperature (\markcite{hae96}Haehnelt
\etal\ 1996), and the range of $U$ parameters reflects this uncertainty as
well as the observed range in \cii/\civ\ ratios.  For pure photoionization
equilibrium $ F  = 3.4\times 10^{-6}\,U^{-1.066}$ and the corresponding
$F$ lies in the range $1.4\times 10^{-4} \to 2.5 \times 10^{-3}$.  As
\markcite{hae96}Haehnelt \etal\ (1996) have emphasized, the temperature
spread caused by shocking in these clouds may result in additional
collisional ionization contributions, which can result in the hydrogen
fraction's being substantially underestimated by photoionization
equilibrium.  However, the $b$ values provide an upper limit to the
temperatures of the clouds since
\begin{equation}
  T \le 10^4 \left ( \displaystyle b \over 13\ {\rm km\ s}^{-1} \right ) ^2
\end{equation}
and we can see from the distribution of $b$ values that there are relatively
few high temperature clouds in the sample.  (At $z=3$ less than 19\% of
the clouds can be above 10$^5$ K.)\ \  We have integrated the collisional
ionization equilibrium though the cloud temperature distribution 
using the assumption that the clouds are thermally broadened to obtain
a mean ratio $N({\rm H})/N$(\hi) = $1.6 \times 10^4$ at $z=3$.  It should
be emphasized that this is an extreme upper limit, since it ignores
blending and kinematic broadening contributions to the $b$ distribution.
The value is smaller at higher redshifts where the $b$ values are lower.

We have no direct measure of the ionization parameter for the lower
column density clouds, and instead must resort to indirect
arguments.  In the linear regime, provided reionization took place early,
$T\sim n^{+0.6}$ and then in ionization equilibrium $\tau_{\rm HI} \sim
n^{+1.6}$ (\markcite{hui96}Hui \etal\ 1996, \markcite{cro97}Croft
\etal\ 1997) which, since the $b$ value -- density distribution is also
relatively invariant as a function of $n_{\rm HI}$, implies $N_{\rm HI}
\sim n^{+1.6}$.  The \hi\ fraction as a function of $N_{\rm HI}$ column
density then scales as
\begin{equation}
	F = {\displaystyle {nT^{-0.7}}\over{J_{21}}} \sim n^{+0.6} 
            \sim N_{\rm HI}{}^{+0.4}.
\end{equation}
In combination with the \hi\ power law indices of $1.38\to1.55$ this then
implies that the contribution to $\Omega_{\rm b}$ is very weakly dominated by
the higher column density clouds just below the break in the DDF, whence
\newcounter{saveeqn}
\newcommand{\alpheqn}{\setcounter{saveeqn}{\value{equation}}%
\stepcounter{saveeqn}\setcounter{equation}{0}%
\renewcommand{\theequation}{\mbox{\arabic{saveeqn}\alph{equation}}}}
\newcommand{\reseteqn}{\setcounter{equation}{\value{saveeqn}}%
\renewcommand{\theequation}{\arabic{equation}}}
\alpheqn
\begin{eqnarray}
    \Omega_{\rm b}  & = & {{H_{\rm o}\over c} {\mu m_{\rm H} \over
    \rho_{\rm c}}} 
    \displaystyle \int_{N_{\rm min}}^{N_{\rm dev}}
    {{f_{\rm dev} (N_{\rm HI}/N_{\rm dev})^{-\beta}\ N_{\rm HI}}
    \over {F_{\rm dev}(N_{\rm HI}/N_{\rm dev})^{0.4}} } \ dN_{\rm HI}\\
  & = & {{H_{\rm o}\over c} {\mu m_{\rm H} \over \rho_{\rm c}}}\,
    \left [ {\displaystyle {f_{\rm dev}\,N_{\rm dev}{}^2} \over
    {(1.6-\beta){F_{\rm dev}} }} \right ] \left [ 1 - \left ( {\displaystyle
    N_{\rm min} \over N_{\rm dev} } \right )^{1.6-\beta} \right ]
\end{eqnarray}
\reseteqn
where $f_{\rm dev}$ is the value of the DDF and $F_{\rm dev}$ the value of
the neutral hydrogen at the $N_{\rm dev}$ column density.  For $\beta=1.45$
and $N_{\rm min}=0.01\,N_{\rm dev}$ we obtain
\begin{equation}
  \Omega_{\rm b} = \left ( \displaystyle {2 \over F_{\rm dev}} \right ) 
       \Omega_{\rm b}({\rm HI})
        = \displaystyle {1.2\times 10^{-7}\,h^{-1} \over F_{\rm dev}} \left ( 
    \displaystyle {N_{\rm dev} \over 2 \times 10^{14}\ {\rm cm}^{-2}}
    \right ) ^{0.5}
\end{equation}
from Table \ref{tbl-4}.  Raising $\beta$ to 1.55, the constant rises to 2.1, 
while reducing $N_{\rm min}$ to 0.001 $N_{\rm dev}$ increases it to 2.43, 
and the combination would increase it to 2.7.

In order to minimize the downwards extrapolation from the metal line
measurements at $N_{\rm HI} \sim 10^{15}\ {\rm cm}^{-2}$ we restrict
ourselves to the highest $z$ range, where $N_{\rm dev}\approx
10^{14.8}\ {\rm cm}^{-2}$ lies at the column density range probed by
the metals.  For $N_{\rm dev}\approx 10^{14.8}\ {\rm cm}^{-2}$, where
$\Omega_{\rm b}$(\hi) = $1.2\times 10^{-7}\,h^{-1}$, and $\beta=1.53$
(Table \ref{tbl-4}), and setting $N_{\rm min} = 10^{11.8}\ {\rm cm}^{-2}$, 
the baryonic contribution from lower column density clouds is then
\begin{equation}
	\Omega_{\rm b} = 0.003\,h^{-1} \left ( {\displaystyle F_{\rm dev} \over
        10^{-4}} \right ) ^{-1}.
\end{equation}
In order to obtain a conservative upper bound we assume that the \cii/\civ\
ratios discussed above correspond to clouds with a
$N($\hi)$\sim10^{16}\ {\rm cm}^{-2}$ and adopt $U=0.03$ at this $N($\hi).
Scaling back to $10^{14.8}$ then gives $F_{\rm dev}$ from photoionization
of $5\times10^{-5}$.  In combination with the upper bound estimate from
collisional ionization, we obtain $F_{\rm dev} \ge 3\times 10^{-5}$ and
$\Omega_{\rm b} \le 0.01\,h^{-1}$.

This is just consistent with low end estimates of the lower bound on
$\Omega_{\rm b}\,h^2$ from \markcite{wei97}Weinberg \etal\ (1997), but
the overlap range is surprisingly small.  This discrepancy can be
directly identified with the adopted limits on $J$ by considering the 
overdensity at $N_{\rm dev}$.  From the definition of $U$
\begin{equation}
	n=2\times10^{-4} \left ( 
          {\displaystyle U \over 0.1} \right ) ^{-1}
          \left ( {\displaystyle J_{21} \over 0.5} \right )\ {\rm cm}^{-3}
\end{equation}
so that the overdensity at ionization parameter $U$ is
\begin{equation}
	{\displaystyle n \over <n>} = 26 \left ( {\displaystyle {\Omega_{\rm b}
         \,h^2} \over 0.01} \right )^{-1} \left ( {\displaystyle 4.5 \over 
         {1+z}} \right )^3 \left ( {\displaystyle U \over 0.1} \right )^{-1} 
         \left ( {\displaystyle J_{21}\over 0.5} \right ).
\end{equation}
Within the models the deviation density corresponds to an overdensity of
roughly 10, and we may rewrite this as
\begin{equation}
	J_{21} = 0.2 \left ( {\displaystyle {n/<n>} \over 10} \right )
                 \left ( {\displaystyle {\Omega_{\rm b} \,h^2} \over 
                 0.01} \right ) \left ( \displaystyle { 1+z \over 4.5} \right )
                 ^3 \left ( {\displaystyle U \over 0.1} \right )
\end{equation}
so that the present results require $J_{21} < 0.2$, which then relaxes the
$\Omega_{\rm b}\,h^2$ constraints in \markcite{rau96}Rauch \etal\ (1996)
and \markcite{wei97}Weinberg \etal\ 1997 at the expense of an inconsistency 
with their lower limit estimates of $J$.

\subsection{The correlation functions} 

The evolution of the correlation strength with redshift is very
ambiguous. As \markcite{mor93}Morris \etal\ (1993) show, the
distribution of the local Ly$\alpha$ does not follow the local galaxy
distribution closely, but neither is it quite random. However, numerical
simulations show that Ly$\alpha$ forest clouds follow the distribution
of dark matter and do show the galaxy distribution more closely at $z
\sim 3$ than does the observed local distribution of Ly$\alpha$ forest
clouds.  As long as the clouds are not randomly distributed, the
evolution of their correlation strength with time provides a glimpse of
structure formation and/or evolution in the early universe.

While the positive correlation over a small velocity range is
marginally significant, another interesting feature, the
anti-correlation, shows contradictory results.  The existence of an
anti-correlation was first claimed for spectra toward Q0055--269
(\markcite{cri95}Cristiani \etal\ 1995) and was later confirmed by
\markcite{mei95}Meiksin \& Bouchet (1995). A similar anti-correlation
is also shown in the correlation function of the Keck HIRES data at
$<z> = 2.85$ (\markcite{hu95}Hu \etal\ 1995). However, with a larger
data set, \markcite{cri97}Cristiani \etal\ (1997) claimed that there
is no anti-correlation. Interestingly, the correlation function
calculated at $<z>=2.85$ for this study (only for Q0014+813 and
Q0302--003) using the same Keck HIRES data does not show the
anti-correlation.  The results of \markcite{hu95}Hu \etal\ (1995)
used 4 QSOs in this redshift interval (the present two together with
Q0636+680 and Q0956+120).  Most of the signal in the anti-correlation
arises from Q0636+680, and when this QSO is excluded the sample does
not show any strong indication of the presence of an
anti-correlation.

This result also gives a possible explanation for the contradictory
results of \markcite{mei95}Meiksin \& Bouchet (1995) and
\markcite{cri97}Cristiani \etal\ (1997).  Since
\markcite{mei95}Meiksin \& Bouchet (1995) used only individual QSOs
to look for an anti-correlation, they were able to find and confirm
the early result of \markcite{cri95}Cristiani \etal\ (1995). However,
with the larger data set, most peculiar features in correlation
functions are averaged, and no longer give the anti-correlation shown
by smaller numbers of QSOs.  This issue requires further work with
much larger data sets.

\section{Summary\label{sec:summary}}

We have analyzed Ly$\alpha$ forest clouds (12.8 $\le \log N_{\rm HI}
\le$ 16.0) towards 5 QSOs in order to study the evolution of these
clouds at $2.2 < z < 3.5$, using Keck{\thinspace}I HIRES data. We
include the Ly$\alpha$ forest cloud analysis at $3.4 < z < 4.0$ given
by \markcite{lu96}Lu \etal\ (1996) from similar quality Keck data to
summarize the evolution of the Ly$\alpha$ forest at $2 < z < 4$.

The differential density distribution function (DDF) does not change
rapidly with redshift at the lower \hi\ column density end, being well
fitted with power-laws close to $\sim 1.4$. However, at the higher
column density end, DDFs depart from a single power-law; at higher
redshift, the amount of deviation from the power-law is smaller and the
deviation column density at which DDFs start to deviate from the
power-law is larger. The number density with redshift including HST
observations at $ z < 2$ indicates mild evolution in the number
density, which reflects the behaviors of the DDFs.  The minimum, mean,
and median line widths of the Ly$\alpha$ clouds tend to increase as the
redshift decreases.  The correlation strength seems to increase as
redshift decreases, at a velocity scale less than $\Delta  v \le
300~{\rm km\ s}^{-1}$. The fluctuation power spectrum at near-Mpc
scales ({\it c.f.}, \markcite{hui96}Hui \etal\ 1996) is found to be
$1.15\pm0.1$, while the contribution to $\Omega_{\rm b}$ from forest
clouds ($12.8 < \log N_{\rm HI} < 14.8$) at $z=3.5$ is $\Omega_{\rm b}
\approxlt 0.01\,h^{-1}$.

\newpage
\begin{table}
\tablenum{1}
\dummytable\label{tbl-1}
\tablenum{2}
\dummytable\label{tbl-2}
\tablenum{3}
\dummytable\label{tbl-3}
\tablenum{4}
\dummytable\label{tbl-4}
\end{table}

\clearpage
%
%
\begin{deluxetable}{lccccccc}
\tablecaption{Fitted QSOs} 
\tablehead{
\colhead{} 	& \colhead{} 	& \colhead{} 	& \colhead{} & 
\colhead{$\lambda_{\rm range}$} & \colhead{} 	& \colhead{} & 
\colhead{} \\[0.5ex]
\colhead{QSO}   & \colhead{V} 	& \colhead{$z_{em}$} & \colhead{$t_{exp}$} & 
\colhead{(\AA)} & \colhead{$z_{{\rm Ly}\alpha}$} & \colhead{$dX^{\rm a}$}
 & \colhead{$\sharp$ of lines}
}
\startdata
  Q1623+268 & 16.0 & 2.521 & $\phn5^{\rm h}\,20^{\rm m}$ & 3850 -- 4195 & 2.17 -- 2.45 & 0.939 & 68{\rlap{$^{\rm b}$}} \nl
  Q1700+643 & 16.1 & 2.722 & $\phn2^{\phm{\rm h}}\,40^{\phm{\rm m}}$ & 3850 -- 4195 & 2.17 -- 2.45 & 0.939 & 82 \nl
  Q0014+813 & 16.5 & 3.387 & $\phn8^{\phm{\rm h}}\,15^{\phm{\rm m}}$ & 4510 -- 4863 & 2.71 -- 3.00 & 1.023 & 118 \nl
  Q0302$-$003 & 17.6 & 3.290 & $10^{\phm{\rm h}}\,00^{\phm{\rm m}}$ & 4510 -- 4863 & 2.71 -- 3.00 & 1.119 & 123 \nl
  Q1422+231 & 16.5{\rlap{$^{\rm c}$}} & 3.620 & $\phn9^{\phm{\rm h}}\,40^{\phm{\rm m}}$ & 5105 -- 5485 & 3.20 -- 3.51 & 1.233 & 146\nl
\enddata
\tablecomments{Magnitudes and redshifts are taken from 
\markcite{ver96}V\'eron-Cetty \& V\'eron (1996).}
\tablenotetext{a}{The redshift path is defined as $X \equiv
0.5 [(1+z)^{2} -1]$, for $q_{\rm o} = 0$.}
\tablenotetext{b}{The number of lines is for $N_{\rm HI} = 10^{13.1} - 10^{14.3}
\ {\rm cm}^{-2}$.}
\tablenotetext{c}{$R$ magnitude.}
\end{deluxetable}
%
%
\begin{deluxetable}{lcccc}
\tablewidth{32pc}
\tablecaption{Incompleteness correction$^{\rm a}$}
\tablehead{%
\colhead{} & \multicolumn{2}{c}{$<z>=2.31$} & \multicolumn{2}{c}{$<z>=3.35$} \\
\cline{2-3} \cline{4-5}\\[-2ex]
\colhead{$\log N_{\rm HI}$} & \colhead{Incompleteness} & \colhead{$\log f$} &
\colhead{Incompleteness} & \colhead{$\log f$}
}
\startdata
  12.8 -- 13.1 & 0.68 & $-11.37$ & 0.44 & $-11.02$ \nl
  13.1 -- 13.4 & 1.00 & $-11.68$ & 0.72 & $-11.54$ \nl
  13.4 -- 13.7 & 0.92 & $-12.04$ & 0.78 & $-11.93$ \nl
  13.7 -- 14.0 & 0.89 & $-12.58$ & 0.89 & $-12.53$ \nl
  14.0 -- 14.3 & 0.90 & $-12.94$ & 0.91 & $-12.90$ \nl
\enddata
\tablenotetext{a}{For the incompleteness correction for the $<z>=2.85$ 
forest clouds, the results of Hu \etal\ (1995) were used.}
\end{deluxetable}
%
%
\begin{deluxetable}{lcc}
\tablewidth{32pc}
\tablecaption{Median $b$ values of the observed QSOs}
\tablehead{
\colhead{} & \colhead{$N_{\rm HI} = 10^{13.8} - 10^{16}\ {\rm cm}^{-2}$} & 
\colhead{$N_{\rm HI} = 10^{13.1} - 10^{14}\ {\rm cm}^{-2}$}\\ 
\colhead{QSO} & (km s$^{-1}$) & (km s$^{-1}$)
}
\startdata
  Q1623+268 & 41  & 36 \nl
  Q1700+643 & 35  & 28 \nl
  Q0014+813 & 34  & 31 \nl
  Q0302$-$003 & 37 & 29 \nl
  Q1422+231 & 30 & 27 \nl
  Q0000-003$^{\rm a}$ & 31 & 29  \nl
\enddata
\tablecomments{Median $b$ values over the indicated column density ranges.}
\tablenotetext{a}{From \markcite{lu96}Lu \etal\ 1996}
\end{deluxetable}
%
%
\begin{deluxetable}{ccccc}
\tablecaption{$\Omega_{\rm b}$ of neutral hydrogen}
\tablehead{
\colhead{Parameters} & \colhead{$<z>=2.31$} & \colhead{$<z>=2.85$} &
\colhead{$<z>=3.35$} & \colhead{$<z>=3.70$}
}
\startdata
  $N_{\rm HI, dev}$ $({\rm cm}^{-2})$ & $10^{14.3}$ & $10^{14.3}$ & $10^{14.8}$ & \omit\hfil---\hfil \nl
  $f(N_{\rm HI})$ & $1.5 \times 10^{6} N_{\rm HI}^{-1.35}$ &
                       $4.9 \times 10^{7} N_{\rm HI}^{-1.46}$ &
                       $4.5 \times 10^{8} N_{\rm HI}^{-1.53}$ &
                       $9.5 \times 10^{8} N_{\rm HI}^{-1.55}$ \nl
  $\Omega_{\rm b}({\rm log}N_{\rm HI} < 14.3)$ & $5.7 \times 10^{-8}$ &
          $6.0 \times 10^{-8}$ & $6.3 \times 10^{-8}$ & $7.2 \times 10^{-8}$\nl
  $\Omega_{\rm b}({\rm log}N_{\rm HI} < 15.5)$ & $1.1 \times 10^{-7}$ & 
          $1.3 \times 10^{-7}$ & $2.1 \times 10^{-7}$ & $2.5 \times 10^{-7}$\nl
\enddata
\end{deluxetable}

\clearpage

\begin{figure}
\figurenum{3b}
\plotone{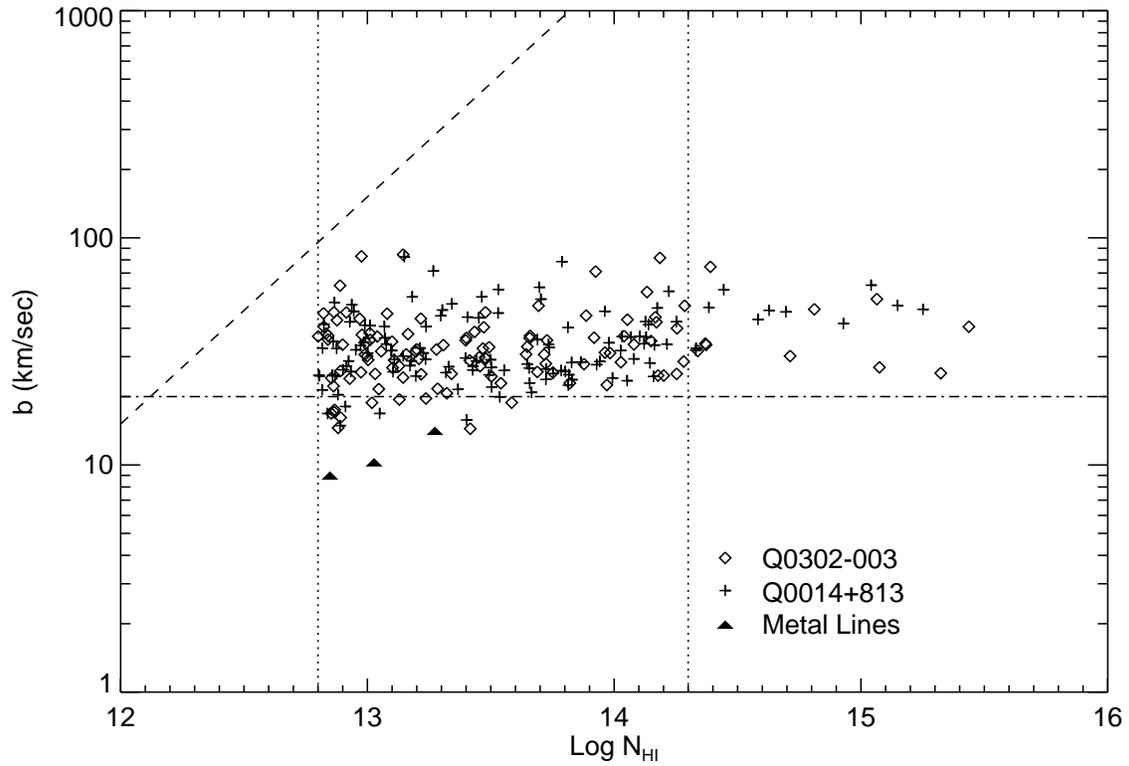}
\caption
{$b - \log N_{\rm HI}$ diagram at $<z>=2.70 - 3.00$.
The dot-dashed line shows $b = 20~{\rm km\ s}^{-1}$. Note that
there are many more lines with $N_{\rm HI} \ge 10^{14}\ {\rm cm}^{-2}$
compared to the same range in column density for $z=2.17 - 2.45$.}
\end{figure}

\begin{figure}
\figurenum{3c}
\plotone{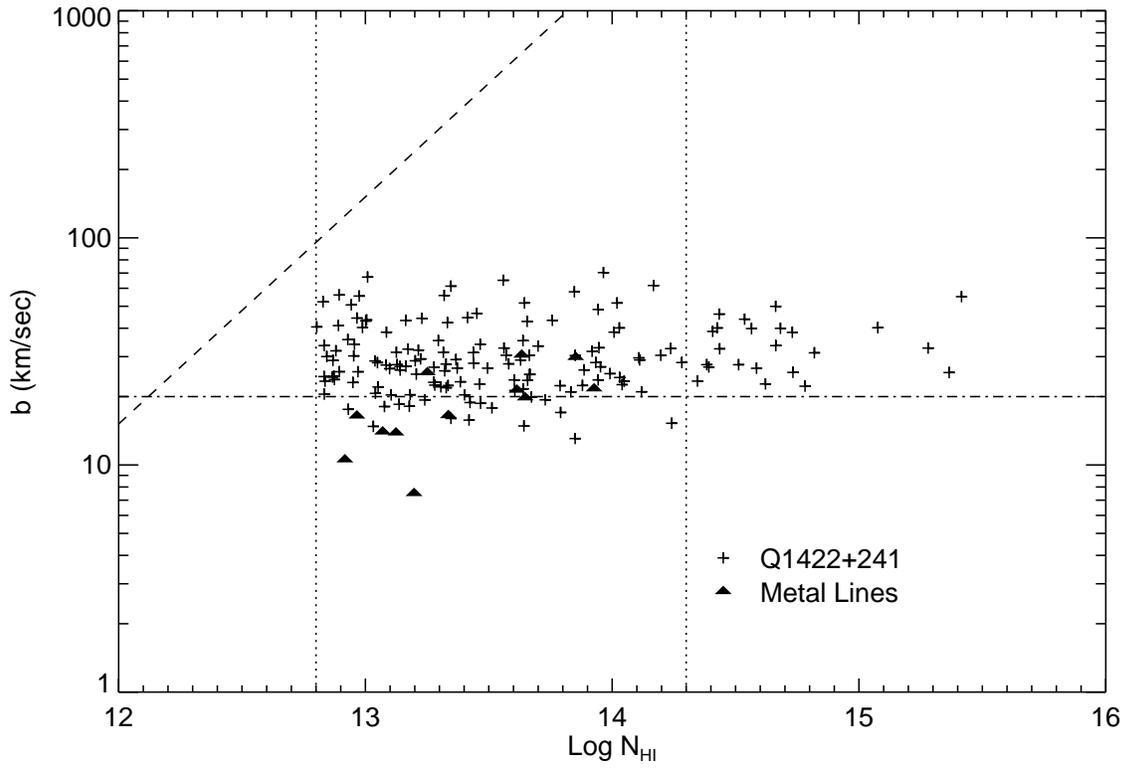}
\caption
{$b - \log N_{\rm HI}$ diagram at $<z>=3.20 - 3.51$.  The
dot-dashed line shows $b = 20~{\rm km\ s}^{-1}$. Note the increase in
the number of lines with $b < 20~{\rm km\ s}^{-1}$ over this redshift
range compared to $<z>=2.31$ and $<z>=2.85$.}
\end{figure}

\begin{figure}
\figurenum{4}
\label{fig:4}
\plotone{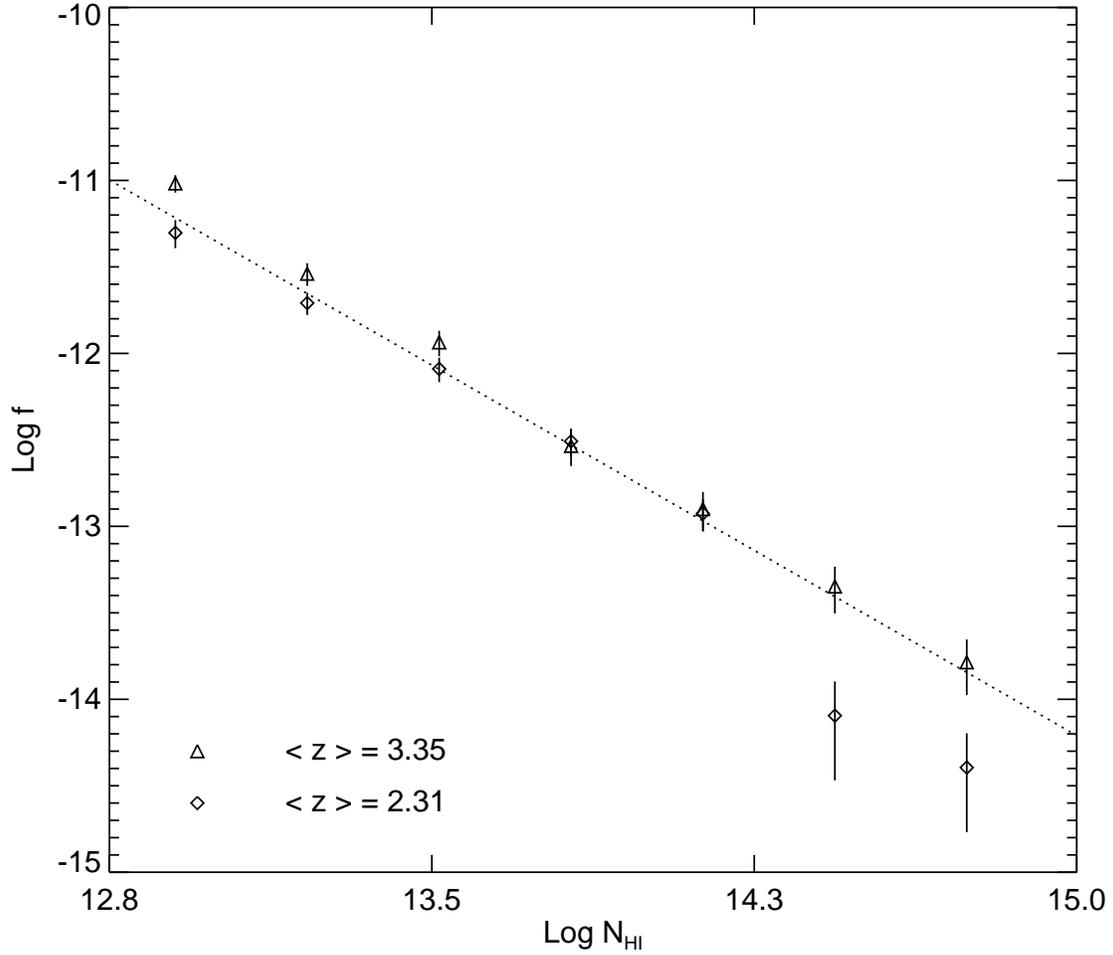}
\figurenum{4a}
\caption
{Incompleteness corrected differential density
distribution functions with redshift. Diamonds are $<z>=2.31$ clouds and
triangles are $<z>=3.35$ systems. $1\sigma$ error bars are shown for each
point.  The dotted line is the same
power-law fit shown in Fig.~\protect{\ref{fig:1}}. See the discussion in 
Sec.~\protect{\ref{subsec:bvalue}} for more details.}
\end{figure}

\begin{figure}
\figurenum{4b} 
\plotone{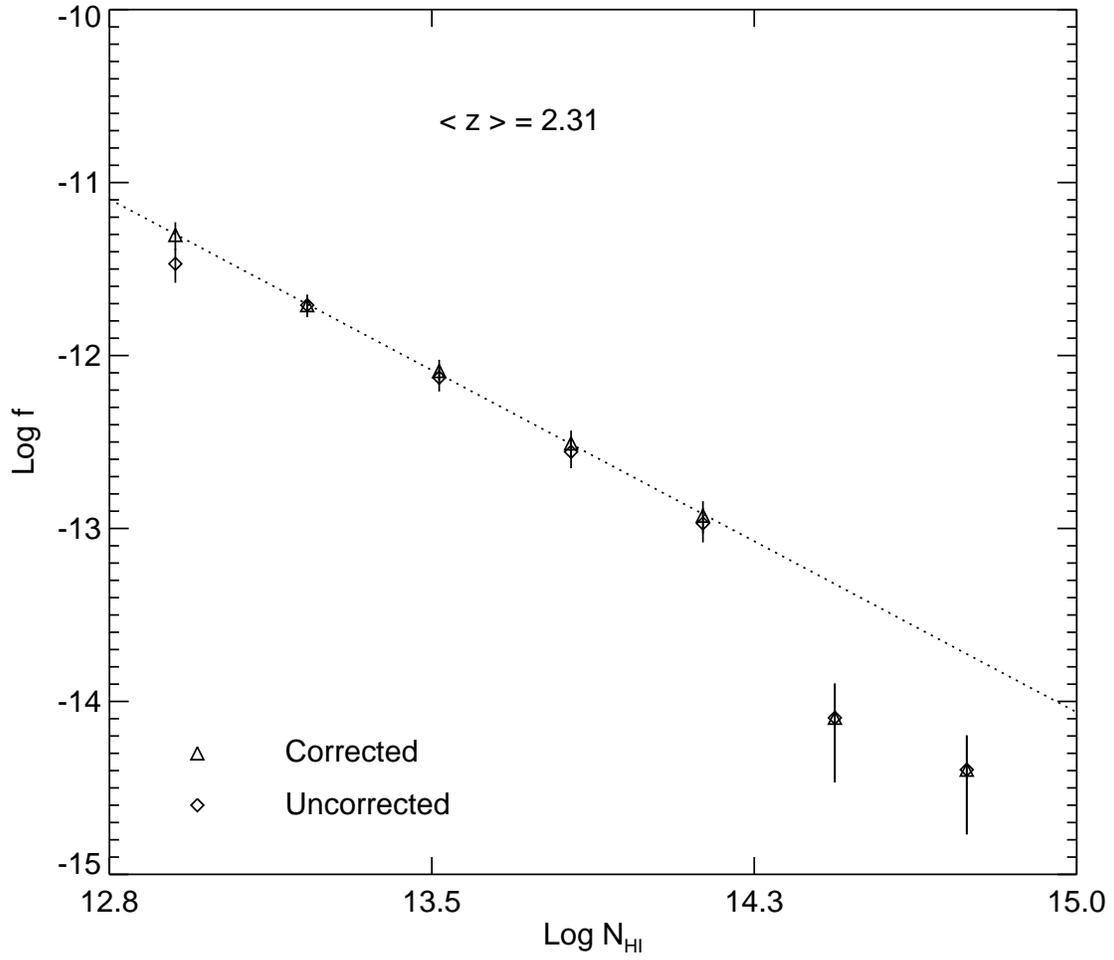}
\caption
{Corrected and
uncorrected differential density distribution functions at $<z>=2.31$.
The deviation from the power-law fit at column densities above
$N_{\rm HI} = 2\times 10^{14}\ {\rm cm}^{-2}$ may be clearly seen.}
\end{figure}

\begin{figure}
\figurenum{4c} 
\plotone{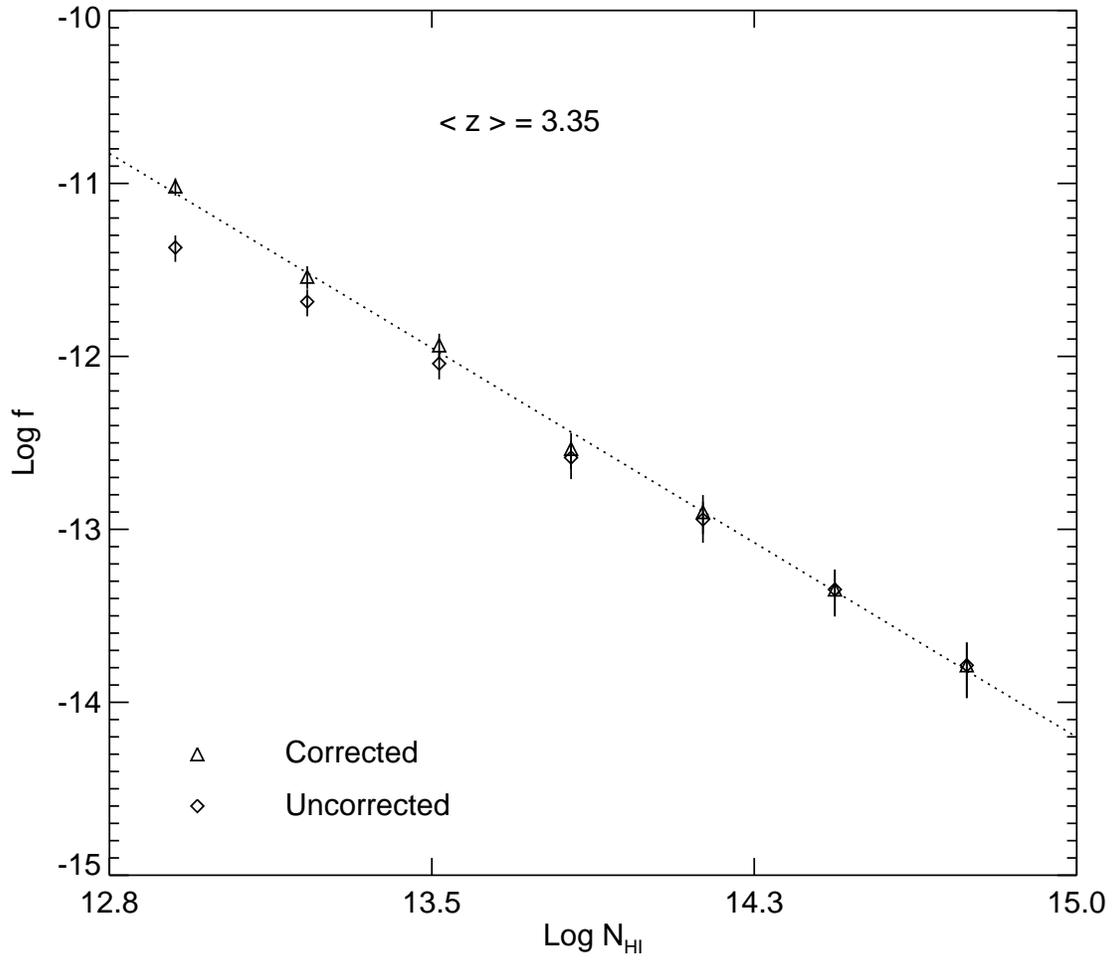}
\caption
{Corrected and
uncorrected differential density distribution functions at $<z>=3.35$.}
\end{figure}

\begin{figure}
\figurenum{5}
\plotone{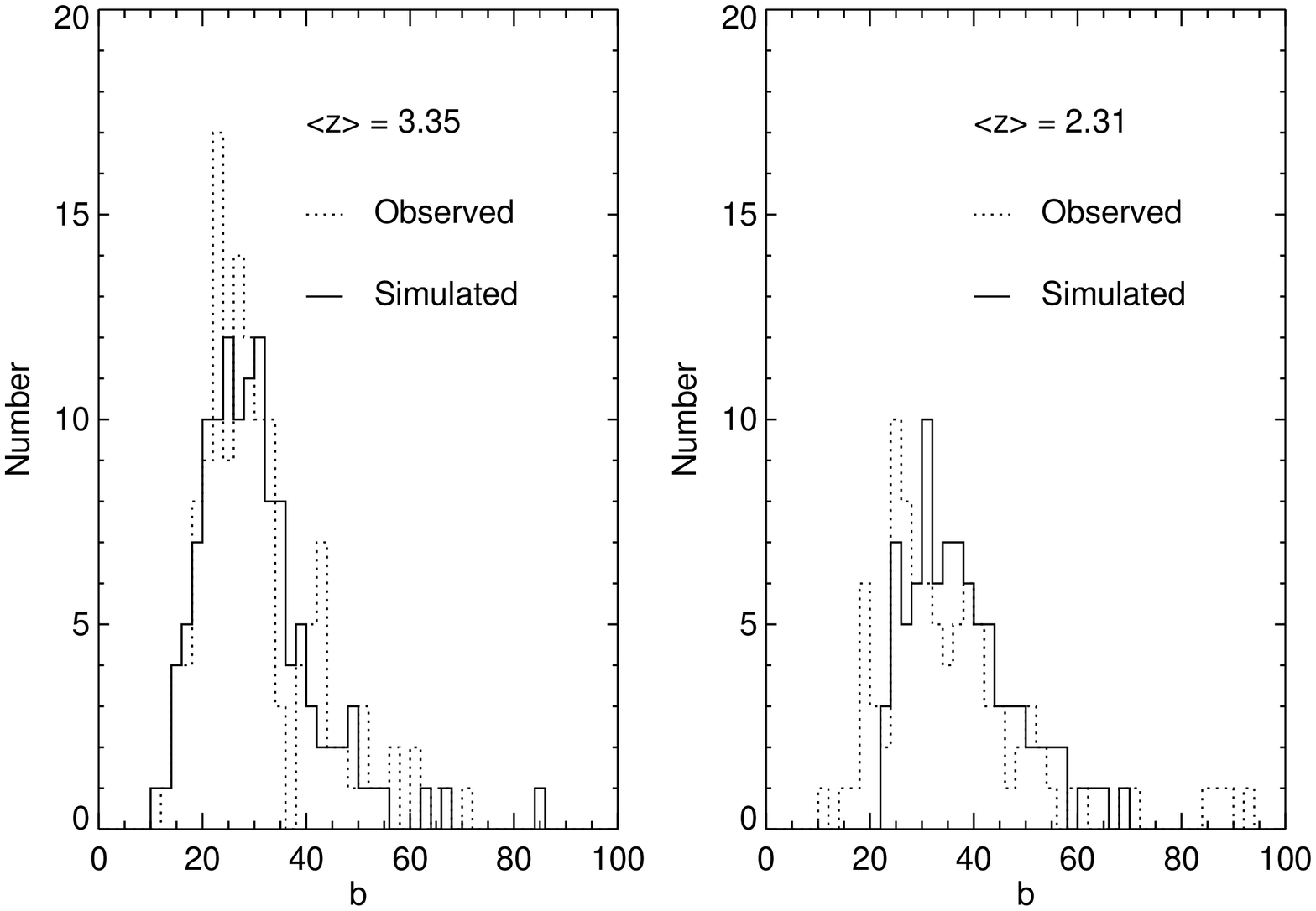}
\caption
{\label{fig:5}Observed (dotted histogram) and simulated (solid histogram)
$b$ distributions. For the
$<z>=3.35$ clouds, the low cut-off $b$ value is $\sim17~{\rm
km\ s}^{-1}$. For the $<z>=2.31$ clouds, there is no single best-fit
assumed distribution for $b$. Assuming that most narrow lines are
unidentified metal lines towards Q1700+643 and ignoring the lines
with $b < 22~{\rm km\ s}^{-1}$, the cut-off $b$ of $24~{\rm
km\ s}^{-1}$ which is shown in the figure best fits the data.}
\end{figure}

\begin{figure}
\figurenum{6a}
\plotone{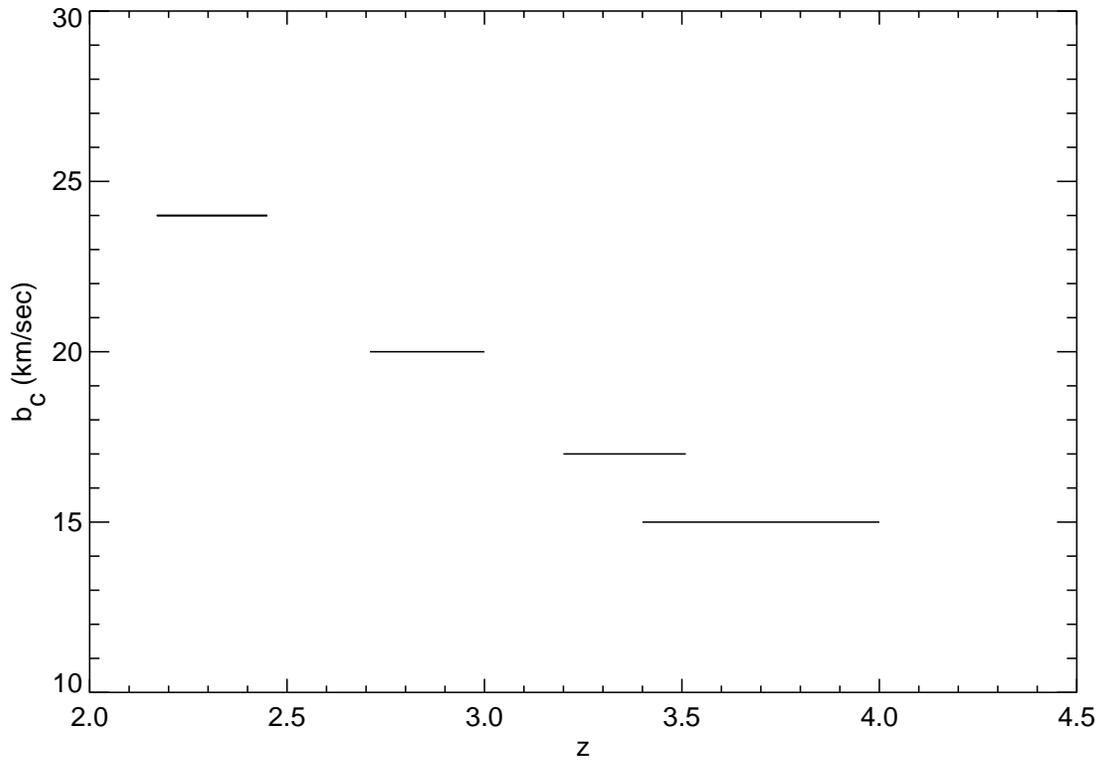}
\caption
{\label{fig:6a}Evolution of $b_{\rm c}$ with redshift.}
\end{figure}

\begin{figure}
\figurenum{6b}
\plotone{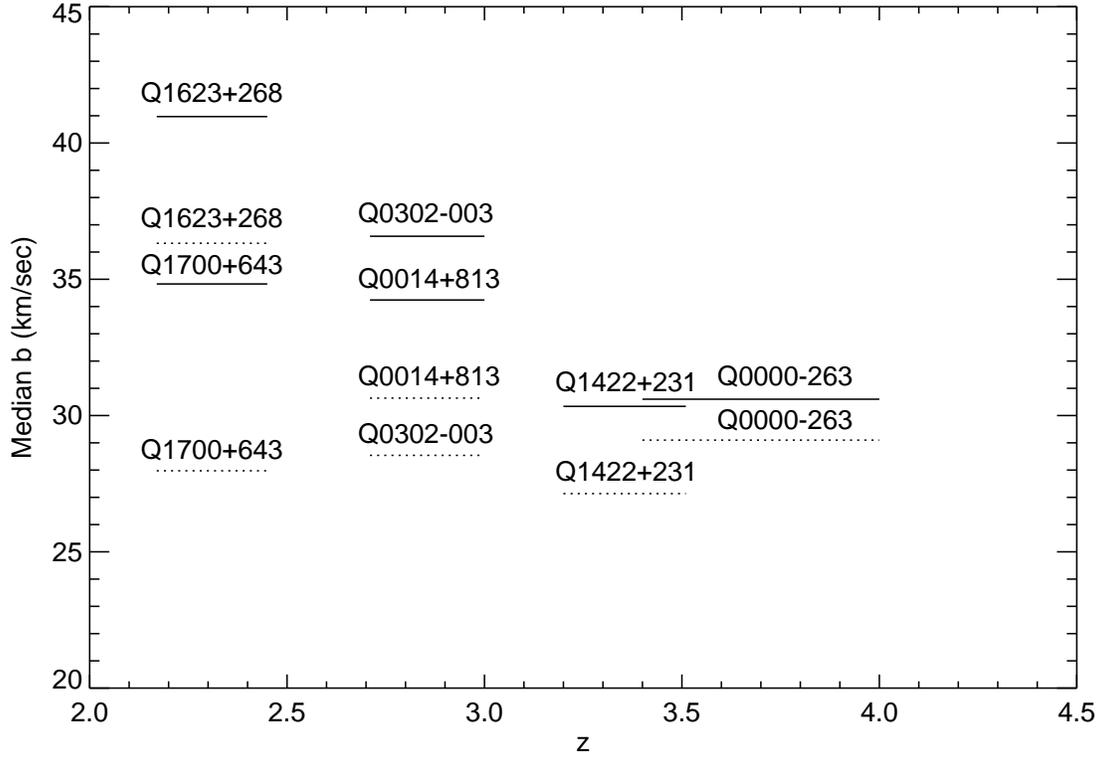}
\caption
{\label{fig:6b}Evolution of median $b$ values with redshift.
The solid line is for $N_{\rm HI} = 10^{13.8} - 10^{16}
\ {\rm cm}^{-2}$, and 
the dotted line is for $N_{\rm HI} = 10^{13.1} - 10^{14}
\ {\rm cm}^{-2}$. There is a general trend of increasing $b$ values
as redshift decreases for both column density selections, though the
effect is more pronounced in the higher column density systems.}
\end{figure}

\begin{figure}
\figurenum{7a}
\plotone{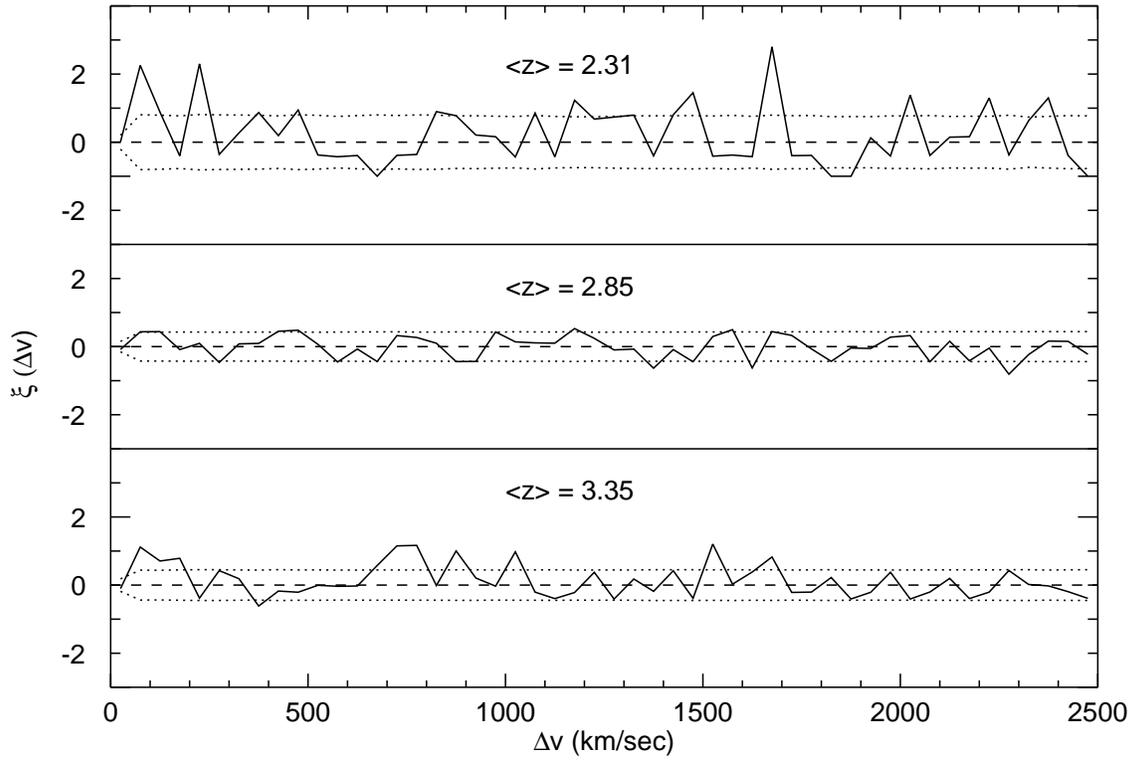}
\caption
{\label{fig:7a}The redshift-averaged correlation functions. The
dotted lines show the 1$\sigma$ Poisson uncertainty. At $<z>=2.31$
and $<z>=3.35$, there is a positive correlation at $\Delta v <
300~{\rm km\ s}^{-1}$.}
\end{figure}

\begin{figure}
\figurenum{7b}
\plotone{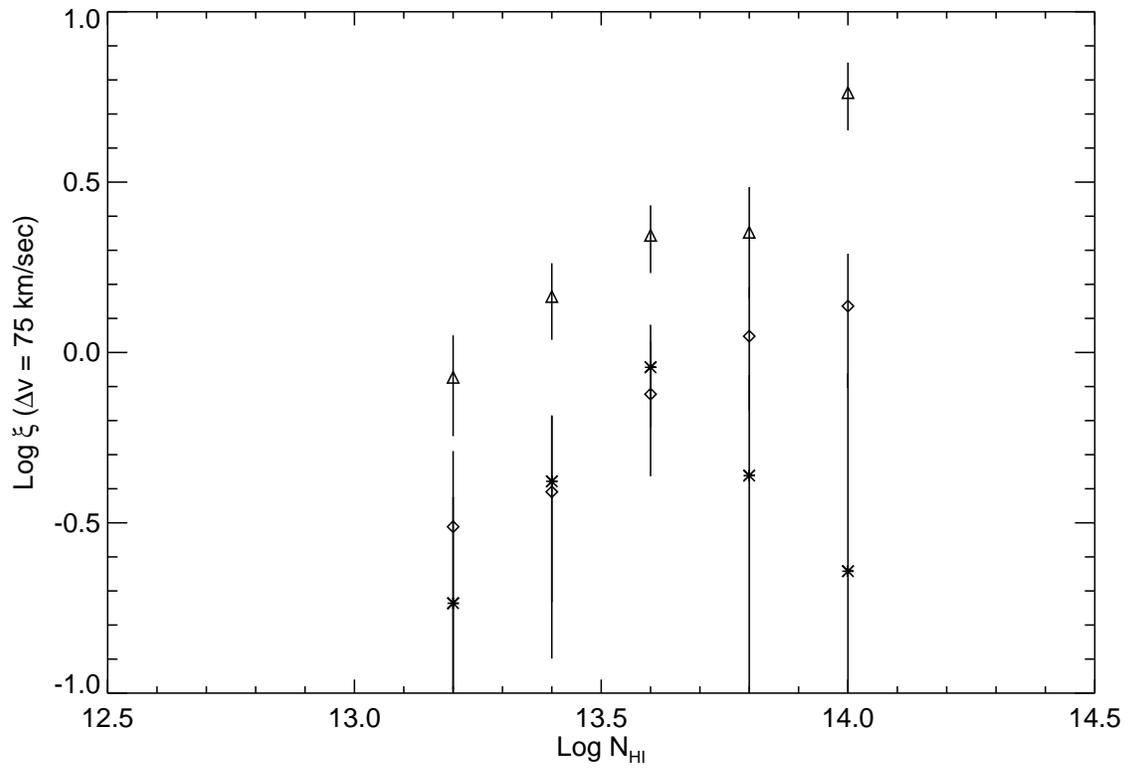}
\caption
{\label{fig:7b}The correlation strengths at $\Delta v =
75~{\rm km\ s}^{-1}$ with $N_{\rm HI, th}$ at different redshifts.}
\end{figure}

\begin{figure}
\figurenum{7c}
\plotone{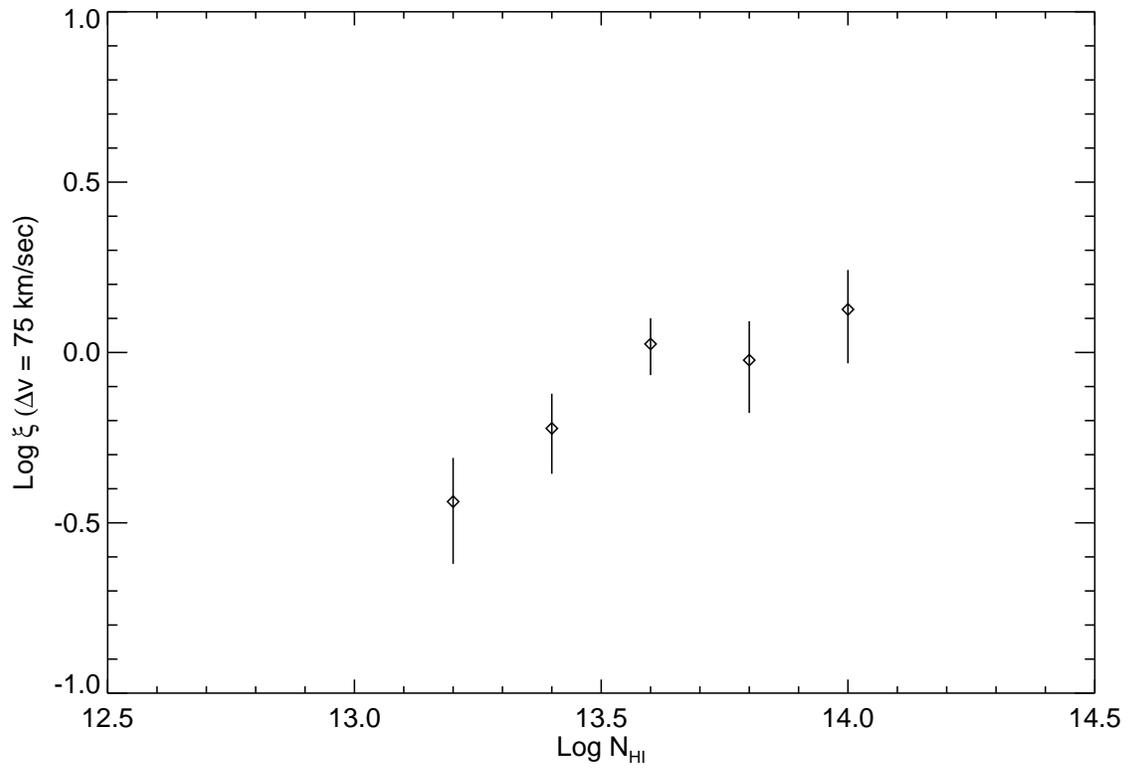}
\caption
{\label{fig:7c}The redshift-averaged correlation strengths at $\Delta v = 
75~{\rm km\ s}^{-1}$ with $N_{\rm HI, th}$.}
\end{figure}

\begin{figure}
\figurenum{8a}
\plotone{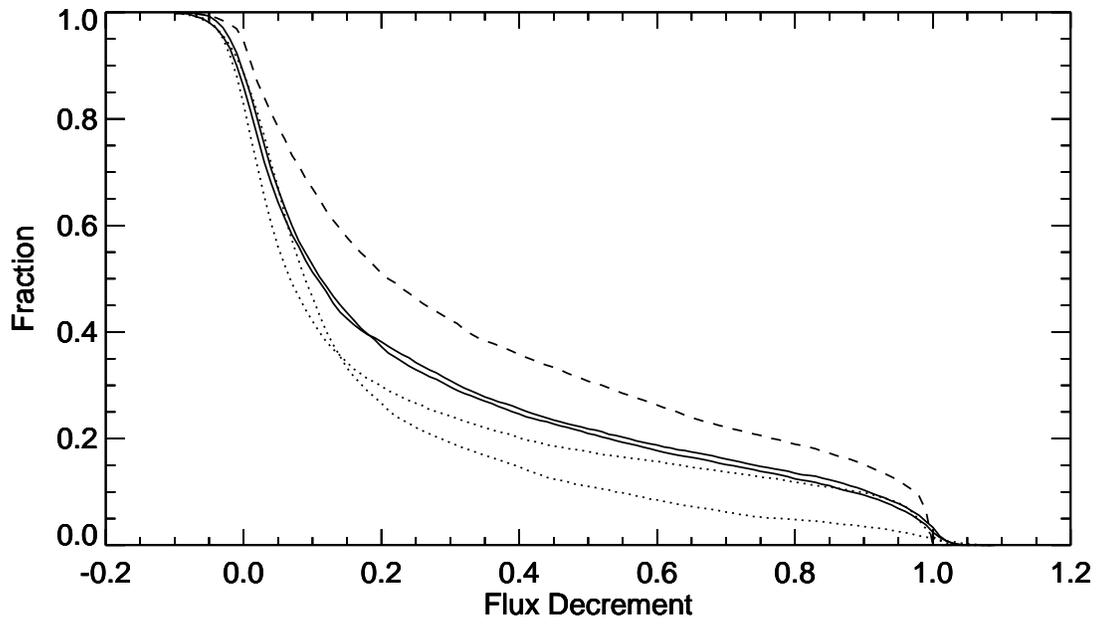}
\caption
{\label{fig:8a}Normalized cumulative distribution of flux decrements
for the 5 quasars. The dashed line shows the $<z>=3.35$ system,
the solid lines the $<z>=2.85$ systems, and the dotted lines the $<z>=2.31$
systems.}
\end{figure}

\begin{figure}
\figurenum{8b}
\plotone{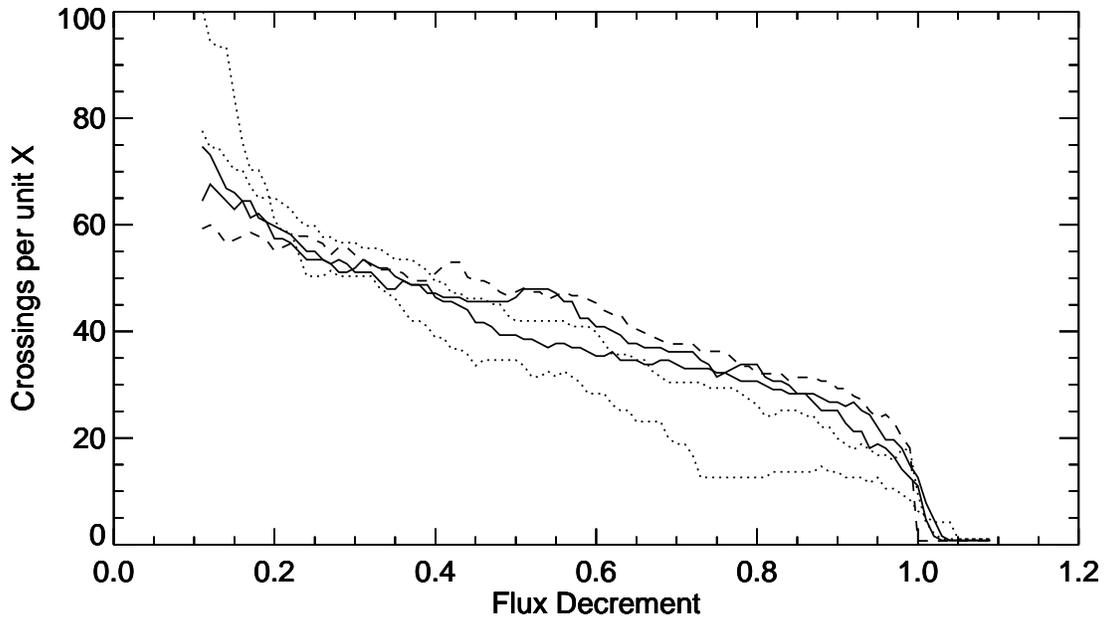}
\caption
{\label{fig:8b}Number of down-crossings per unit X plotted vs. flux decrement.
Redshift identifications as for part a).}
\end{figure}

\begin{figure}
\figurenum{9}
\plotone{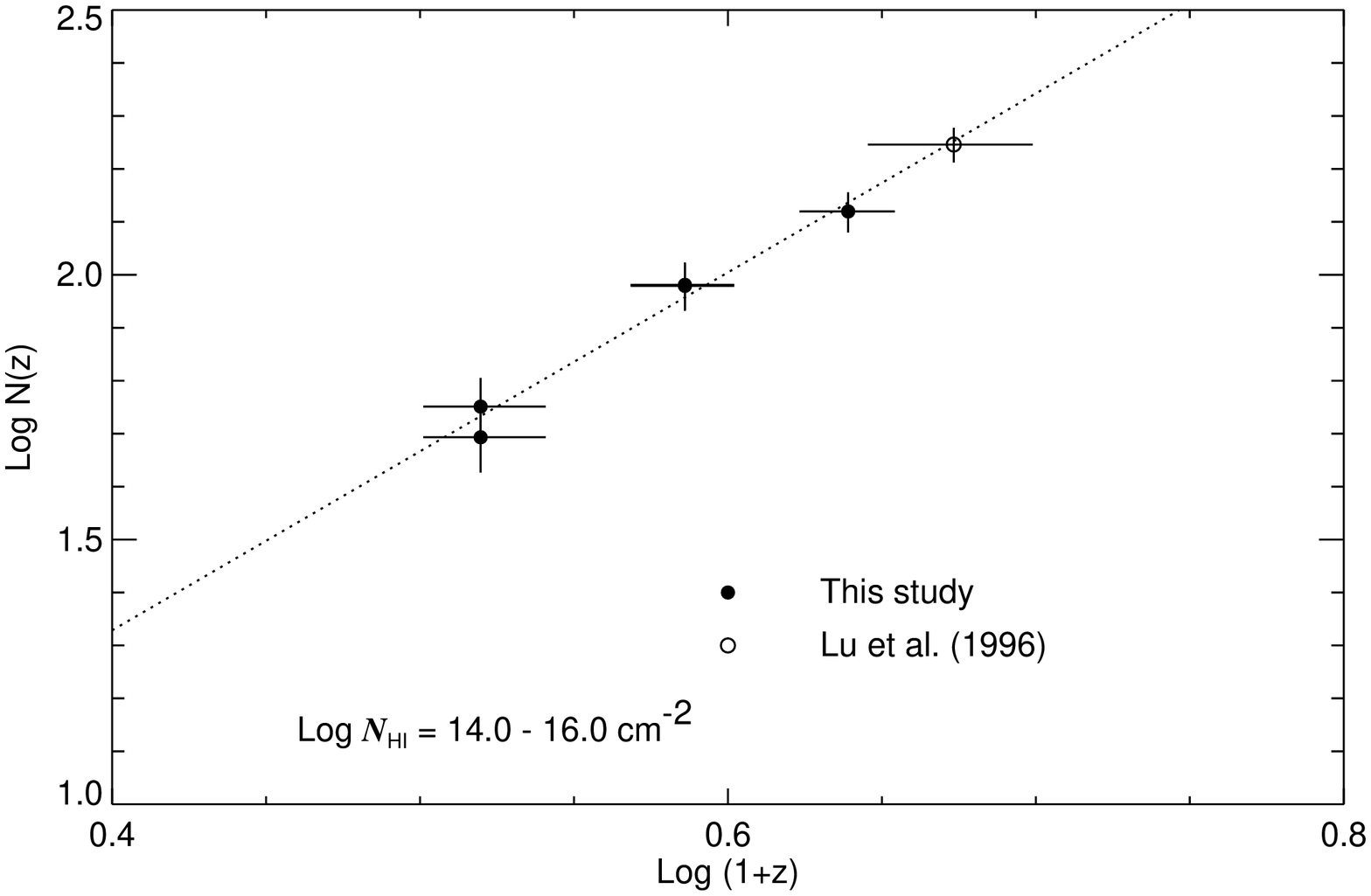}
\caption
{\label{fig:9}$N(z)$ vs redshift at $N_{\rm HI} = 10^{14-16}
\ {\rm cm}^{-2}$.  The filled circles show our data points, while the
open circle is taken from \protect{\markcite{lu96}}Lu
\etal\ (1996).  For Q0014+813 and Q0302--003, the $N(z)$ values are nearly
identical, and are marked with a single overlapping data point.  
The data show a uniform trend with a steep rise in the number
density evolution.  The dotted line shows a least-squares fit
to $N(z)\propto (1+z)^{3.38}$.}
\end{figure}

\begin{figure}
\figurenum{10}
\plotone{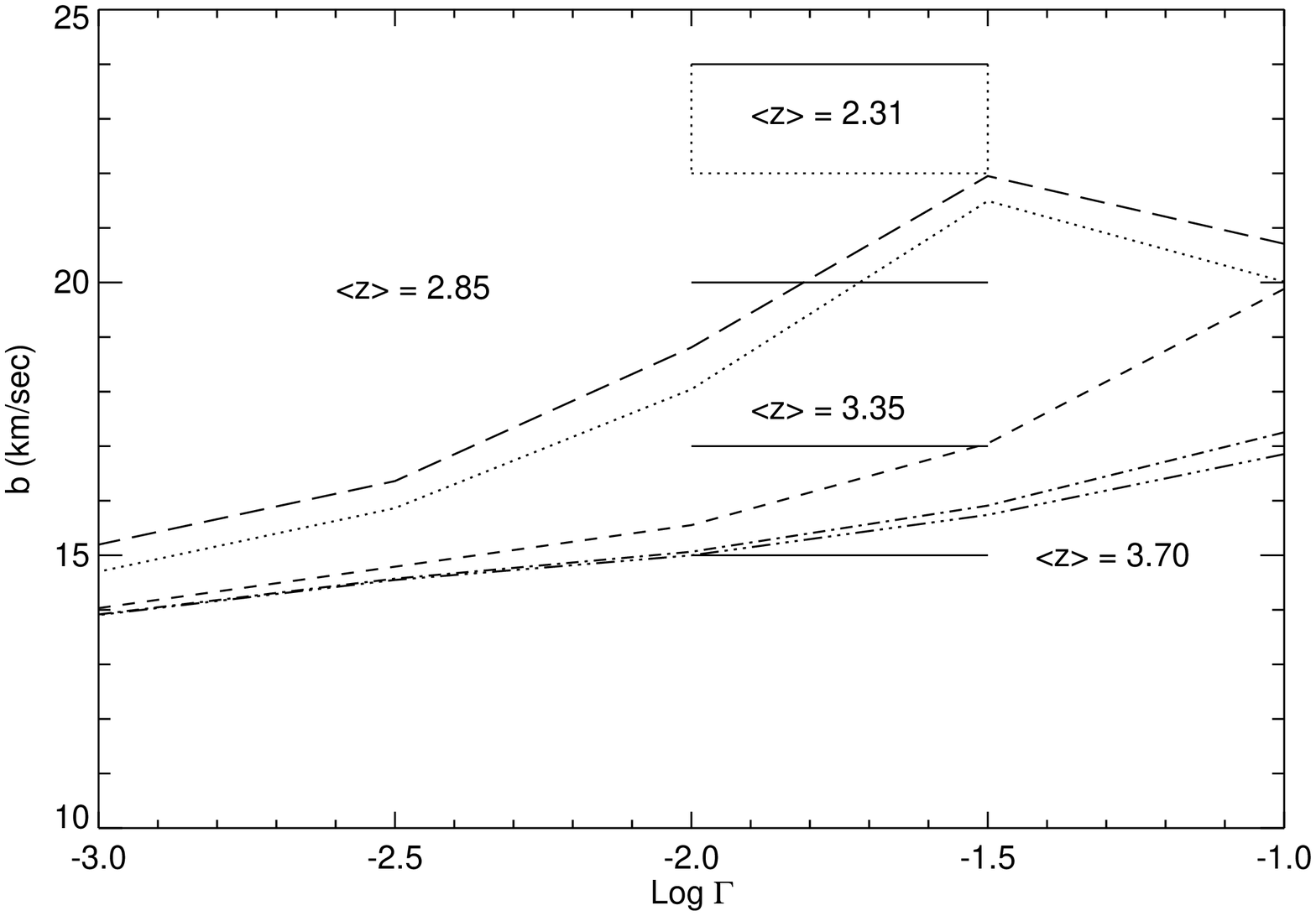}
\caption
{\label{fig:10}Photoionization and $b$ values. 
The lines were calculated from the photoionization model for the
assumed \heii\ ionization at $z \sim 3.1$. See the discussion in
Sec.~\protect{\ref{subsec:bevol}} for more details.}
\end{figure}

\end{document}